# ChatGPT as Co-Advisor in Scientific Initiation: Action Research with Project-Based Learning in Elementary Education

Fabiano Villan  a
Renato P. dos Santos  b

ª Universidade Luterana do Brasil (ULBRA), Curso de Licenciatura em Física
ᵇ Universidade Luterana do Brasil (ULBRA), PPG em Ensino de Ciências e Matemática (PPGECIM),

## ABSTRACT

**Background**: In the contemporary educational landscape, technology has the power to drive innovative pedagogical practices. Overcoming the resistance of teachers and students to adopting new methods and technologies is a challenge that needs to be addressed. **Objectives**: To evaluate the effectiveness of ChatGPT as a co-advisor in research projects and its influence on the implementation of Project-Based Learning (PBL), as well as overcoming resistance to the use of new pedagogical methodologies. **Design**: An action-research methodology was employed, including unstructured interviews and the application of questionnaires via Google Forms. **Setting and Participants**: The research was conducted in an elementary school, involving 353 students and 16 teachers. **Data Collection and Analysis**: Data were gathered through observations and notes in meetings and interviews, complemented by electronic questionnaires, with quantitative and qualitative analyses performed via Microsoft Excel and Google Forms. **Results**: The introduction of ChatGPT as a pedagogical tool led to increased student engagement and decreased teacher resistance, reflected in recognition at local science fairs. **Conclusion**: The study confirmed the utility of ChatGPT in school research co-orientation, highlighting its role in facilitating PBL and promoting cultural changes in educational practice, with proactive school management identified as a catalysing element in adapting to educational innovations.

_______________________
Corresponding author: Renato P. dos Santos. Email: RenatoPSantos@ulbra.edu.br

**ChatGPT como Coorientador na Iniciação Científica: Pesquisa-ação com Aprendizagem Baseada em Projetos no Ensino Fundamental**

## RESUMO

**Contexto**: No cenário educacional contemporâneo, a tecnologia tem o poder de impulsionar práticas pedagógicas inovadoras. A resistência de professores e alunos em adotar novos métodos e tecnologias é um desafio a ser superado. **Objetivos:** Avaliar a eficácia do ChatGPT como coorientador em projetos de pesquisa e sua influência na implementação da Aprendizagem Baseada em Projetos (ABP), bem como na superação da resistência ao uso de novas metodologias pedagógicas. **Design:** Utilizou-se uma metodologia de pesquisa-ação, incluindo entrevistas não estruturadas e aplicação de questionários via *Google Forms*. **Cenário e Participantes:** A pesquisa foi realizada em uma escola de ensino fundamental, envolvendo 353 alunos e 16 professores. **Coleta e análise de dados:** Os dados foram coletados através de observações e anotações em reuniões e entrevistas, complementadas por questionários eletrônicos, com análises quantitativas e qualitativas realizadas via *Microsoft Excel* e *Google Forms*. **Resultados:** A introdução do ChatGPT como ferramenta pedagógica resultou em maior engajamento dos alunos e diminuição da resistência dos professores, refletida em reconhecimento em feiras científicas locais. **Conclusão:** O estudo confirmou a utilidade do ChatGPT na coorientação em pesquisa escolar, evidenciando seu papel em facilitar a ABP e promover mudanças culturais na prática educativa, destacando a gestão escolar proativa como um elemento catalisador na adaptação a inovações educacionais.
**Palavras-chave:** ChatGPT como coorientador, Iniciação Científica, Ensino Fundamental, Aprendizagem Baseada em Projetos, Pesquisa-ação.

## INTRODUCTION

In the contemporary educational landscape, characterized by a rising demand for integrated practical and theoretical skills, challenges emerge to adopt innovative approaches that enrich students' learning journey (Aureliano & Queiroz, 2023). Pedagogical evolution, driven by technological innovation, becomes an urgent necessity to meet the demands of the 21st century (Rodrigues, 2021). To meet this demand, educators must become flexible and innovative professionals, capable of integrating active methodologies with technologies, to provide transformative and meaningful learning experiences (Aureliano & Queiroz, 2023).

In this context, Paulo Freire emphasized that

> "There is no teaching without learning, the two explain each other, and their subjects, despite the differences that denote them, are not reduced to the condition of object, one of the other. Those who teach learn in teaching, and those who learn teach in learning." (Freire, 1996, p. 25)

For this author, this continuous interaction reinforces the central role of the educator in the learning journey.

In this panorama, Generative Artificial Intelligence (GenAI) emerges, a category of AI that, unlike merely analysing data, creates original content, such as texts or images, based on what it has learned. Well-known examples of GenAI applications include ChatGPT and BingChat, among others, which appear as promising educational allies, broadening thematic horizons and deepening the learning process.

ChatGPT, when integrated with the Problem-Based Learning (PBL) methodology, has the potential to enrich the educational experience, aligning with Vygotsky's ideas (1991), expanding students' Zone of Proximal Development (ZPD) and contributing to the professional development of teachers.

Thus, an innovative approach is proposed to promote meaningful and interdisciplinary learning, by using ChatGPT to explore contemporary themes that can go beyond the traditional training of educators.

However, many educators show resistance to adopting these technologies. The reluctance is often grounded in the perception that such innovations may stray from the true educational purpose or in the hesitation to abandon traditional practices. Therefore, understanding this resistance becomes the central problem of this research.

The aim of this study is to analyse the impact of ChatGPT as a co-advisor in elementary school science fair projects. Through the lens of Vygotsky's ideas about the facilitator role of the educator, it seeks to understand how the conscious incorporation of advanced technologies can shape teaching practices and influence the quality of contemporary education, while exploring the challenges and resistances faced by educators in integrating new technologies into their practices.

## THEORETICAL FRAMEWORK

The education of children and youth in the 21st century should prioritize the development of essential skills for academic, professional, and personal success. Creativity and socio-emotional skills are vital in this context. Creative individuals tend to be open to new experiences, imaginative, and unafraid to express ideas, even in the face of complex information (de Cássia Nakano, Primi, & Alves, 2021). Scientific Initiation (SI) not only acquaints students with research methods and practices but also incites critical skills, curiosity, and autonomy (National Academies of Sciences, Engineering, and Medicine, 1997). Thus, SI can play a crucial role in stimulating these skills, preparing students for the challenges of the 21st century (da Silveira, Cassiani, & Von Linsingen, 2021).

In Brazil, SI in school contexts, encouraged by the state, had a significant milestone in 2003 with the establishment of the Junior Scientific Initiation Program (PIC-Jr) by the National Council for Scientific and Technological Development (CNPq). Before this period, similar initiatives were already implemented in some educational institutions, notably the Scientific Vocation Program (PROVOC) of the Oswaldo Cruz Foundation (Fiocruz) since 1986. Notably, the Brazilian government program was largely influenced by the Fiocruz model, adopting principles aimed at cultivating talents for science and encouraging scientific vocations (Brazil, 2015).

SI in Elementary Education is fundamental in students' educational formation. Thus, SI has the potential to serve as a foundation for the development of future researchers, paving the way for more advanced learnings and the emergence of scientific vocations. Participation in a Scientific Initiation Program (PIC) promotes the development of a theoretical-scientific attitude, clarity in vocational choices, research skills, and the expansion of study group performance with the goal of increasing scientific productivity (Nascimento & Morosini, 2019). A study by Saliba et al. (2019) observed that doctoral students who underwent scientific initiation obtained better scientific production, achieving a more significant number of publications compared to those without this experience.

In the SI process, the pedagogical mediation performed by the teacher is vital. Carvalho, Nevado, and Menezes (2007) emphasize that an effective advisor must go beyond simply instructing; they should inspire their students to actively engage in research. This involves encouraging them to reflect on their findings and to explore, rather than just seeking ready answers. This idea is encapsulated in the statement by Hernández and Ventura (1998): "Not everything can be taught through projects, but everything can be taught as a project."

Explaining how learning is enhanced by mediation and collaboration, Vygotsky's theory remains central to this research. According to his idea of the Zone of Proximal Development (ZPD), a student/learner might be ready to acquire a new knowledge or skill, yet may still be unable to do so independently:

> The zone of proximal development defines those functions that have not yet matured but are in the process of maturation, functions that will mature but are currently in an embryonic state. These functions could be called 'buds' or 'flowers' of development rather than 'fruits' of development. The actual level of development characterizes mental development retrospectively, while the zone of proximal development characterizes it prospectively. (Vygotsky, 1991, p. 97).

Vygotsky emphasized the crucial role of the social environment and interaction in the construction of knowledge, introducing concepts such as the ZPD to explain how learning is enhanced by mediation and collaboration. The learning process according to the ZPD is marked by the difference between a student's ability to act independently and their ability to act with adequate support (Vygotsky, 2001).

On the other hand, the theoretical foundation of the Problem-Based Learning (PBL) methodology is often associated with educators like Dewey, who highlighted the relationship between practice and learning, Bruner, who advocated for new educational proposals, as well as constructivists like Piaget and Vygotsky (Servant-Miklos et al., 2019). Its current form emerged as an educational innovation in 1969 at McMaster University in Canada, influenced by earlier experiences at Harvard, with the goal of promoting self-directed learning and critical thinking (Servant-Miklos et al., 2019), with the idea that learning would be more engaging if the learner were actively involved in their own learning process (Schmidt, 2012).

This approach not only emphasizes the teacher's role as facilitator and mediator but also fosters the development of cognitive skills such as observing, questioning, analysing, and problem-solving, stimulating students' curiosity, creativity, and critical thinking, as advocated in General Competence 2 of the Brazilian Education Guidelines and Bases (Brazil, 2018).

Furthermore, this pedagogical mediation enables the development of social skills, such as communication abilities, teamwork, leadership, cooperation, and social responsibility, as described in General Competence 3, preparing students to be active citizens aware of their role in society (Brazil, 2018).

The action-research methodology, initially proposed by Kurt Lewin, stands out as a practical and participatory approach, aimed at solving problems and promoting social and educational changes through a cyclical process of planning, action, and reflection (Lewin, 1946). This method emphasizes collaboration between researchers and participants, with the goal of transforming the reality in question. Following this principle, Kemmis and McTaggart (1988) developed a structured model that guides educators and researchers in the planning and execution of action-research projects, proposing a reflective sequence that involves the continuous reconstruction of understanding and practice. This methodology is particularly relevant in the field of education, where the dynamics of teaching and learning can be enhanced by the direct and reflective application of new pedagogical strategies, allowing educators and students to co-construct knowledge in a collaborative and contextualized manner.

On the other hand, since the early experiments with Artificial Intelligence (AI) in the 1950s, technology has been shaping the educational landscape. The transition into the 1980s and 1990s saw the emergence of pioneering AI tools dedicated to education (Mekari, 2023). Cognitive models like the Student, alongside platforms such as Khan Academy, revolutionized teaching, providing more intuitive interactions and personalized feedback for students. The beginning of the 2000s brought a qualitative leap forward. Adaptive learning platforms emerged, and giants like Google, Microsoft, and Apple introduced innovative tools. Among them, Google Classroom stood out, transforming the dynamics between students and teachers by promoting individualized teaching, adapted to each student's pace.

ChatGPT, a notable member of the Large Language Models (LLMs) family from OpenAI, is a tangible example of this integration. Its creation involved training on large volumes of textual data, resulting in its ability to respond to a wide range of questions and commands in natural language. ChatGPT's capacity to process natural language and generate contextualized responses enables it to become an effective co-advisor, assisting students and teachers on their educational journeys.

However, the resistance and insecurity of faculty in teaching degree programs concerning Digital Information and Communication Technologies (DICT), including AI applications like ChatGPT, are still evident. This hesitation, evidenced by the prevalence of traditional teaching methods, has both technical and emotional roots. To overcome such barriers, it is essential to provide teacher training that integrates affective, technical, and pedagogical dimensions, enabling the full exploration of these technologies' potential and promoting a pedagogical transformation adapted to the digital world (Silva & Paniago, 2022).

Parallelly, teaching methodologies have also evolved, and the STEM (Science, Technology, Engineering, and Mathematics) approach has gained prominence, promoting interdisciplinary learning and problem-solving orientation. Educational robotics emerged as a valuable pedagogical tool, stimulating cognitive abilities, creativity, and innovation.

In the contemporary educational context, Vygotsky's Zone of Proximal Development (ZPD) intertwines with technological innovations, particularly with AI. While pedagogical mediation has evolved from its traditional format, positioning the educator as essential in knowledge construction, AI has established itself as a vital tool for amplifying learning.

In summary, when considering the current educational landscape, it becomes clear that the intersection of Vygotsky's ZPD with technological innovations, mainly AI, is redefining teaching. Pedagogical mediation, which has traditionally been centred around the educator, now benefits from these technologies. Teachers become not just facilitators but active mediators, creating enriched and dynamic learning experiences. The combination of innovative pedagogical methods with AI facilitates deeper learning, equipping students for the challenges of the 21st century. This theoretical framework will be used to explore this interaction more deeply in the subsequent section.

## THE ZPD AND CHATGPT AS A CO-ADVISOR IN SCIENTIFIC GUIDANCE

In the contemporary educational landscape, Vygotsky's Zone of Proximal Development (ZPD) has generated significant discussions when combined with technological innovations such as ChatGPT to enhance learning and evolve pedagogical mediation. It is therefore essential to observe the practical manifestation of this combination in the contemporary educational setting, with ChatGPT serving as a co-advisor in scientific guidance. Over the years, pedagogical mediation has undergone a metamorphosis, with technology increasingly playing a role in enhancing learning. When introduced as a co-advisor, ChatGPT becomes an embodiment of this evolution, acting as a technological extension of the human educational mediator and offering students perspectives and insights that transcend their current capabilities. This combination has the potential to broaden the ZPD for both students, allowing them to tackle more advanced concepts with appropriate support from teachers, and teachers themselves, enabling them to address more complex topics than they could independently.

In the educational context, ChatGPT acts not just as a mere virtual assistant but as a technological extension of the human educational mediator, assisting both students and teachers in expanding their horizons. It can clarify complex concepts, provide insights for research, and more importantly, collaborate in the co-advising of learning projects. Such integration resonates with the fourth General Competence of the National Common Curricular Base (BNCC) of Brazil, which emphasizes the adoption of technologies in education (Brazil, 2018).

However, despite significant advancements in Artificial Intelligence in Education (AIEd) providing educators with a range of new tools, there is growing concern about their impacts. The need for an ethical approach becomes evident when considering the increasing interaction between humans and AI in the educational context. While there are ethical guidelines for AI, guidance specifically aimed at basic education in Brazil is still in its infancy. According to Adams et al. (2023), analyses of recent guidelines revealed indispensable ethical principles such as Transparency and Pedagogical Suitability. The focus of the ethical use of these AI tools should be on the accuracy of information and the development of students' critical thinking.

Yet, it is crucial to understand that the advent of ChatGPT does not diminish the central role of the educator. Instead, it serves to enhance and complement teaching, aligning pedagogy with Vygotsky's principles of social and mediated learning. AI serves as a tool for technological mediation, not as a substitute for the teacher. Teachers play a critical role in guiding and interpreting the information provided by ChatGPT, tailoring it to the specific needs of the students. This promotes deeper and more meaningful learning.

A study by Borges and Borges (2018) on the practice of teachers as advisors in scientific guidance in basic education reports various challenges. Indeed, the process of teaching and guidance requires from the teacher specific knowledge and competencies, which are often unconscious but indispensable for successful teaching. These skills are mobilized during the teaching and guidance process and are crucial for Science Education (Gil-Pérez & Carvalho, 2001; Borges, Rita, & Vitérico, 2022).

Thus, AI applications like ChatGPT, when used appropriately, acting as technological mediators, allow students to advance beyond their individual abilities, expanding their ZPD. Through interaction with ChatGPT, students can approach more challenging topics, explore new perspectives, and expand their understanding.

Santos et al. (2020) discussed the difficulties faced by teachers in attempting to implement active methodologies such as Problem-Based Learning (PBL) in public schools. Two of the main obstacles identified are the lack of time for teachers to plan their actions using these differentiated methodologies and the need to guide students on topics beyond their expertise.

When asked, ChatGPT itself highlights some of the reasons why teachers may feel uncomfortable with this pedagogical practice (Figure 1).

**Figure 1**

*Initial dialogue exploring challenges in scientific guidance in primary education.*

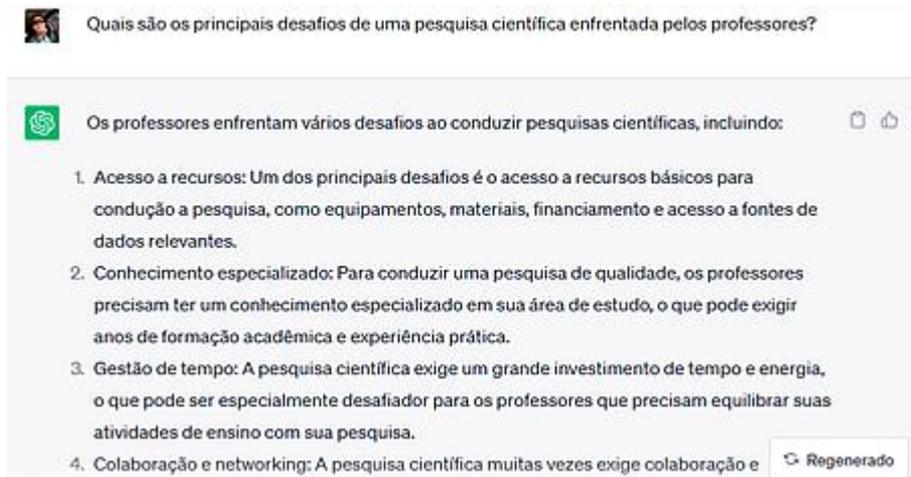

Translation Note:

The primary challenges faced by teachers when conducting scientific research include:

1. Access to resources: One of the main challenges is access to basic resources for conducting research, such as equipment, materials, funding, and access to relevant data sources.
2. Specialized knowledge: To conduct quality research, teachers need specialized knowledge in their area of study, which can demand years of academic training and practical experience.
3. Time management: Scientific research requires a significant investment of time and energy, which can be particularly challenging for teachers who need to balance their teaching duties with their research.
4. Collaboration and networking: Scientific research often requires collaboration and networking, which necessitates efficient communication and coordination among colleagues.

By integrating ChatGPT with Vygotsky's ZPD, a dynamic platform is created where students are guided toward deeper levels of understanding. However, this approach not only benefits the students. It also reflects on the development of the teacher, resulting in a symbiosis of learning.

Nevertheless, it is notable that in some situations, teachers manifest resistance to guiding research groups that exceed the limits of their training area. This challenge is further heightened when discussing the incorporation of emerging technologies, such as ChatGPT, into education.

However, amid this discussion, the figure of the teacher as a technological mediator emerges as a vital bridge between technology and student learning. This teaching role emphasizes Vygotsky's ideas, reinforcing the importance of pedagogical support in merging AI with educational practice. By adopting and integrating technology as a co-advisor, the teacher not only overcomes initial barriers but also enhances their teaching methodology. The ability to learn to learn, intrinsic to the teaching profession, assumes a central role in this context. This learning to learn is, therefore, also a *learning by doing*, in John Dewey's classic formulation of pedagogy.

This perspective is also supported by Perrenoud (2000), who emphasizes the need to work through problems and projects, proposing complex tasks and challenges that incite students to mobilize their knowledge and, to some extent, complete it. This presupposes an active, cooperative pedagogy, open to the city or neighbourhood, whether in urban or rural areas. Teachers must stop thinking that teaching is the core of the profession. Teaching today should consist of designing, fitting, and regulating learning situations following active and constructivist pedagogical principles.

By embracing technology as a co-advising partner, the teacher demonstrates an openness to constant improvement and professional development. This ability not only reflects the essential adaptability of modern educators but also fosters the idea that education is a joint journey, where teachers and students evolve together.

The collaborative interaction and pedagogical mediation, central to Vygotsky's ideas, take on a new dimension in the Problem-Based Learning (PBL) approach with the aid of ChatGPT in guiding independent study. As students work in teams to solve complex issues, ChatGPT takes on the role of co-advisor, offering insights and encouraging the teacher and students to explore beyond their individual capacities, expanding their understanding through interaction with technology.

In summary, at the confluence of these approaches, an environment emerges where pedagogical mediation, both by the teacher and by ChatGPT, becomes essential for the active construction of knowledge. Vygotsky's principles, when incorporated into the PBL methodology and interaction with ChatGPT, not only guide the investigative journey but also highlight the importance of the teacher's intentionality in creating a space for discovery, collaboration, resilience, and mutual growth.

## METHODOLOGY

The study was carried out at the Municipal School of Primary Education IP in Canoas, RS, which caters to both the early and final years of primary education. After nearly eight years without a science fair, the school management was motivated to revisit and update their pedagogical approaches.

Inspired by Vygotsky's theories on social interaction in learning and the expansion of the ZPD, we adopted the PBL methodology to investigate the implementation of innovative pedagogical practices, with a particular focus on the potential use of ChatGPT as a co-advisor in undergraduate research guidance.

As Vygotsky emphasized, innovative approaches to scientific problems often require the creation of entirely new methods of investigation and analysis, which go beyond mere modifications of previously accepted methods (Vygotsky, 1991).

The choice of PBL and action research for this study is based on their complementarity and effectiveness for the educational context in question. PBL places students in the position of active problem-solvers, promoting the development of essential critical and interdisciplinary skills for scientific initiation. Simultaneously, action research serves as a reflective and adaptive mechanism for educators, allowing for continuous improvement of pedagogical practices through iterative cycles of planning, action, observation, and reflection. This synergy between the two approaches not only facilitates more meaningful and contextualized learning for students but also promotes a dynamic and responsive teaching practice, capable of adjusting to emerging needs and the challenges of contemporary teaching. Thus, the joint implementation of PBL and action research in this study represents a deliberate effort to align educational theory with innovative and effective practices, ensuring that learning and development occur in an integrated and sustainable manner.

It is vital to highlight that the flexibility of teaching practice and the promotion of interdisciplinarity are cornerstones of PBL, and action research served as a complementary tool to guide teaching decisions and strategies. The active involvement of the teacher in the scientific initiation process was essential, as highlighted by Carvalho, Nevado, and Menezes (2007). This involvement includes the use of questions intended to unsettle the certainties built by students, thereby encouraging a deeper reflection on the content.

In line with the foundations of PBL, each research project in Scientific Initiation was conceived as a complex problem that required students to adopt an investigative and reflective approach. This pedagogical model placed students at the centre of the learning process, encouraging them to identify and define research questions, search and analyse relevant information, formulate hypotheses, and develop strategies for data collection and analysis.

Such an approach promoted not only the acquisition of specific knowledge on the chosen topic but also the development of fundamental competencies such as critical thinking, problem-solving, and communication skills. In this way, learning occurred organically and meaningfully, through direct engagement of students with their research projects, thus aligning with the goal of PBL to integrate theory and practice in rich and contextualized educational experiences.

Due to some discomfort from the lead teachers in guiding research outside their areas of expertise, the majority of teachers decided, in agreement with the management team, that the research topics should align with each lead teacher's training, which partially diverted the IC process from the essence of PBL.

In this scenario, the researcher, with the assistance of ChatGPT, was available to guide groups that opted for an interdisciplinary approach, maintaining the principles of PBL and testing the effectiveness of ChatGPT as a co-advisor.

The assistance of ChatGPT, acting as a co-advisor, was indispensable for the researcher to effectively guide groups in an interdisciplinary approach, expanding the scope beyond their own academic training. This digital co-advisor was essential for providing insights on a diverse range of student research topics, deepening scientific questioning, organizing information, and supporting strategic decisions. Its application extended from literature review to the exploration of new pedagogical approaches, offering a rich and diverse perspective that goes beyond the traditional limits of academic guidance.

This innovative use of ChatGPT demonstrates its applicability as a pedagogical tool, capable of complementing and enriching the guidance process. Furthermore, the flexibility and adaptability of AI allowed the researcher and students to explore interdisciplinary themes, promoting a richer and more diversified learning experience.

The researcher also played an active role in the development and refinement of teaching materials and research strategies adopted by the groups, in partnership with the school and lead teachers (Figure 2).

Throughout the entire process, ChatGPT served as a complementary tool utilized by the researcher, contributing to the detailed review of the proposal and suggested themes, as well as optimizing the execution of the planning for guidance in elementary SI, resulting in tangible, clear, and effective improvements. The ethical and transparent use of ChatGPT strengthened the collaboration between the researcher and the tool, demonstrating its value in the self-training process for educators. The support offered by the researcher helped to mitigate common concerns of teachers, such as lack of time and the challenges of guidance, facilitating a more efficient and effective pedagogical practice.

On the other hand, action research played a crucial role in this educational intervention, serving both as a methodology and reflective practice. Through it, the researcher not only implemented PBL (Problem-Based Learning) and ChatGPT as teaching and learning tools but also actively participated in the educational process, collaborating directly with students and teachers. This hands-on approach allowed for dynamic adjustments and immediate responses to the emerging needs of the school environment.

**Figure 2**

*Educational material developed to clarify the scientific initiation process and the scientific method.*

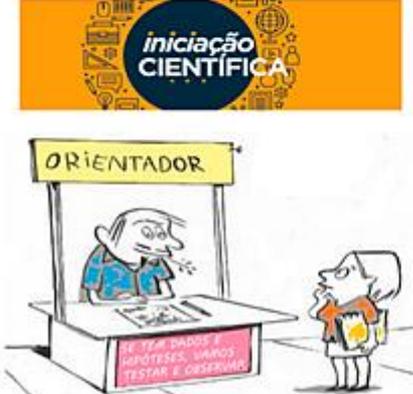

Translation note (Only main text translated):

Orientation to Scientific Initiation in Elementary Education

By Municipal Elementary School Irmão Pedro.

The Scientific Method: definitions, applications, main stages

Scientific research is essential for technological progress and for solving widespread problems in today's society. From a simple curiosity to complex scientific investigations, the use of a systematic method, the scientific method, is imperative. See the main stages of the scientific method:

THE FUNDAMENTAL STAGES OF THE PROCESS

- Observation
- Questions
- Hypotheses
- Experiment
- Analysis
- Conclusion

Initially, face-to-face interviews were conducted individually with 353 students from the final years of elementary school, covering classes from the 6th to the 9th grade, both morning and afternoon sessions, one at a time, to assess their prior knowledge about the scientific method. During these interviews, the researcher recorded detailed observations in a logbook. Concurrently, structured questionnaires via Google Forms were administered to the 16 head teachers, who were guiding the research groups, to collect information about their academic qualifications, pedagogical practices, and comfort with guiding SI projects, detailed in Appendix A. Similarly, an additional survey via Google Forms was conducted with the students from the groups guided by the researcher to measure their receptivity to SI, with the results presented in Appendix B[1].

---

[1] This study did not involve the submission of a research project to a formal ethical evaluation by competent boards, as the activities carried out were an integral part of a teaching internship, being inserted in the everyday educational context. Similarly, there was no application of the Informed Consent Form, since data collection was performed during routine teaching internship activities, aligned with standard pedagogical practices. However, full responsibility for any consequences arising from the study is assumed, committing to assistance and reimbursement for any damages to the research participants, in accordance with Resolution No. 510 of April 7, 2016, by the National Health Council of Brazil, absolving the journal Acta Scientiae from any implications resulting from this work.

Data collection was carried out following a mixed-methods approach. The responses from the questionnaires, automatically compiled and analysed by Google Forms, were reviewed to provide pertinent quantitative and qualitative insights into the study. Quantitative analysis was performed using Microsoft Excel, with the creation of pie charts to visually illustrate the percentage distributions, facilitating the understanding of the results and highlighting significant trends. This mixed methodology ensured a holistic understanding of the phenomena under study, allowing for the correlation of the researcher's direct observations with the responses obtained from the questionnaires.

The data collected allowed for an ongoing assessment of the research groups, enriched by the employment of the action research methodology, and served as a basis for strategic decisions by the school administration, leading to the overcoming of initial challenges with confidence and paving the way for more effective and innovative practices.

The school adopted a fundamental strategy to strengthen guidance and SI of research groups, through the organization of strategic pedagogical stops. During these times, the teachers of each class could reserve class periods exclusively to direct the orientation of the involved students.

During this process, teachers and students were educated on the scientific method in face-to-face meetings twice a week. To support the teachers, the school authorized the creation of a virtual environment (Figure 3) where they could seek information about the scientific method. This facilitated access to digital resources, space for teacher training, and allowed teachers to become familiar with the proposed pedagogical approach.

This structured approach ensured that teachers had adequate time and support to effectively conduct guidance with students, a crucial component in the implementation of the new pedagogical approach.

This strategy not only valued direct interaction between teachers and students but also provided a conducive environment for in-depth discussions, clarification of doubts, and the promotion of constructive dialogue.

The project concluded in two distinct stages. Initially, all school groups presented their research work following the scientific method, guided within the teaching practice aligned with the PBL methodology. These presentations underwent a rigorous assessment aimed at analysing and ranking the works according to the rubrics established in the SI project notice. As a result of this assessment, 30 of the 102 workgroups were selected for the science fair that took place in July 2023 at the school, involving the entire school community.

**Figure 3**

*Virtual environment developed to support teachers and students with materials on the scientific method.[2]*

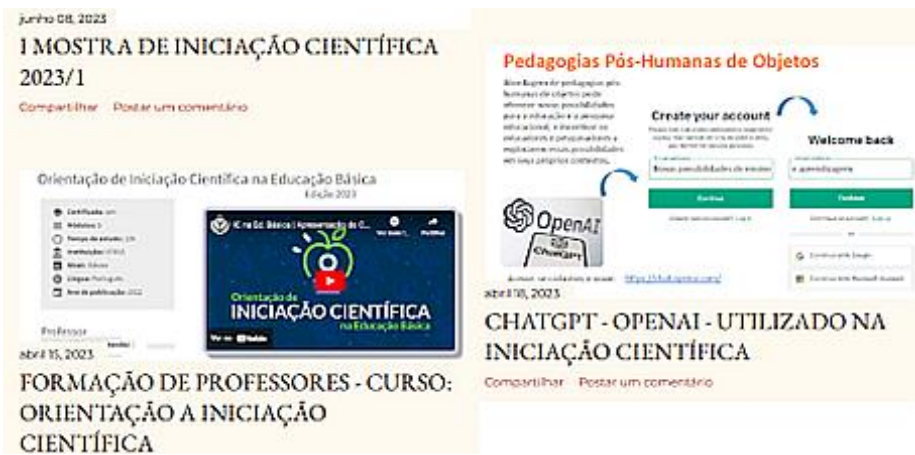

Translation note (Only main text translated):

1st SHOWCASE OF SCIENTIFIC INITIATION 2023/1

Orientation to Scientific Initiation in Biblical Education

TRAINING OF TEACHERS - COURSE: ORIENTATION TO SCIENTIFIC INITIATION

Post-Human Pedagogies of Objects

CHATGPT - OPENAI - USED IN SCIENTIFIC INITIATION

The evaluation was carried out by external evaluators and the school's teachers; at this point, the researcher intentionally stepped back, not participating in the evaluations considering the adopted action-research approach. Here, the importance of the teacher reflecting on their pedagogical practice with an open mind is highlighted, not only with the goal of professional growth but also to benefit the students. The school recognized the efforts of the teachers and students with certificates.

---

[2] https://iniciacaocientificaemefip.blogspot.com/

## RESULTS AND ANALYSIS

This section summarizes the results and analyses of the impact of innovative teaching-learning methodologies on the scientific initiation of students. Initially, the study investigates the students' prior knowledge of the scientific method and their willingness to engage in scientific initiation, as well as assessing the resistance of teachers to new pedagogical practices and the challenges in implementing the project. Subsequently, the research examines the variety of research themes chosen by the students and performance at the final showcase, including a brief statistical study on the influence of the type of guidance on the outcomes. Finally, the co-guidance by ChatGPT is described, with specific examples, highlighting its integration into enhancing research projects and the added value to the students' educational experience.

### Initial Perceptions and Barriers to Educational Innovation

Firstly, the initial data collection from students revealed valuable information (Figure 4), such as the fact that 70.8% of students use Google as a research tool, while only 3.4% seek guidance from their teachers.

**Figure 4**

*Analysis of research sources used by students.*

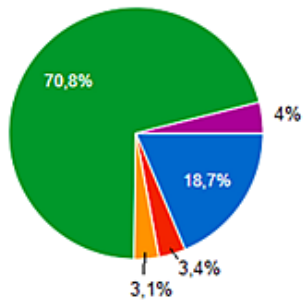

2 - Quando tu quer saber algo ou tirar alguma dúvida, qual seu principal meio de pesquisa?

353 respostas

- a) Normalmente pergunto aos meus pais
- b) Normalmente pergunto aos meus profs
- c) Pesquiso na Wikipédia
- d) Pesquiso sempre no google
- e) Pesquiso sempre no youtube
- f) Livros, revistas

70,8%   4%   18,7%   3,4%   3,1%

Translation note:

2. When you want to know something or clear a doubt, what is your main method of research?

353 responses

a) I usually ask my parents
b) I usually ask my teachers
c) I always search on Wikipedia
d) I always search on Google
e) I always search on YouTube
f) Books, magazines

The results of the survey conducted during the first guidance sessions of the work groups to assess student receptivity, analysed using *Google Forms*, indicated that 68% of the students are predisposed to participate in the scientific initiation proposal, showing motivation and interest (Figure 5). However, the results also show that 32% of the students indicated some form of unwillingness, among whom, students from the 6th and 9th grades have the highest rates of rejection.

**Figure 5**

*Survey of student receptivity to the scientific initiation proposal - Percentage of students predisposed and unwilling.*

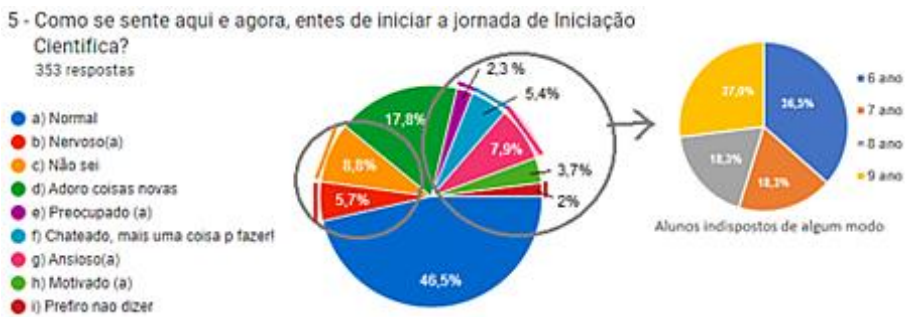

Translation note:

5. How do you feel right now, before starting the journey of Scientific Initiation?

353 responses

a) Normal
b) Nervous
c) I don't know
d) Excited
e) Looking for new things
f) Worried
g) Anxious, another thing I must do!
h) Motivated
i) I prefer not to say

On the other hand, the questionnaire administered to the 16 lead teachers, responsible for guiding the research groups, revealed little prior experience among the faculty on the use of the PBL methodology. Only 75% responded to the questionnaire, and we found that 66.7% of these teachers have no prior experience with PBL teaching methodology (Figure 6). These data highlight the difficulties encountered in applying the methodology and reinforce the need to provide clear guidelines and support to teachers.

**Figure 6**

*Teachers' experience in scientific initiation guidance and PBL methodology - Percentage of teachers with prior experience.*

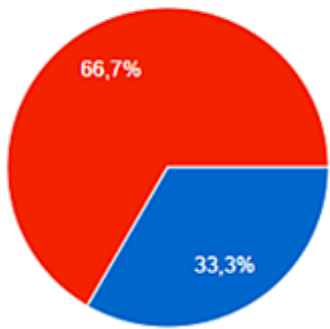

8 - Já trabalhou com aprendizagem baseada em projetos na escola ?

12 respostas

● Sim
● Não

66,7%

33,3%

Translation note:

8. Have you ever worked with project-based learning at school?

12 responses

Yes - 33.3%
No - 66.7%

Seeking a first reflection with the faculty about the difficulties in implementing the proposed methodology (Figure 7), other concerns stand out among the teachers such as allocating the necessary time for guidance (58.6%) and discomfort with guiding themes outside their area of expertise (25%).

**Figure 7**

*Difficulties in implementing the proposed methodology - Teachers' concerns.*

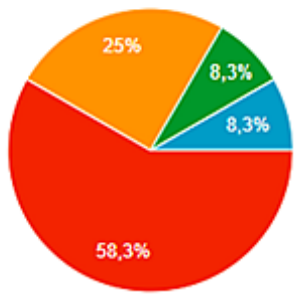

Translation note:

10. What is the greatest difficulty you believe can be encountered in the orientation process for scientific initiation that we are starting?

12 responses

Blue: Lack of interest from students
Red: Time management
Orange: Guidance on themes outside my area of training
Green: Little collaboration from families
Purple: Poor collaboration from other colleagues/peers
Light Blue: Lack of institutional support
Pink: Questions related to ethics regarding student participation in research
Light Green: Lack of financial resources of students

In a second reflection with teachers about the challenges in implementing the project (Figure 8), the underestimation of students by the faculty stands out, with 50% believing that the lack of student interest could be a difficulty. While the concern about guiding students on topics outside their area of expertise adds up to 41.7%. These challenges indicate the need to work on building teachers' confidence in their students, recognizing and valuing their capabilities, in addition to providing support in guiding the chosen themes.

**Figure 8**

*Challenges in implementing the project - Teachers' concerns.*

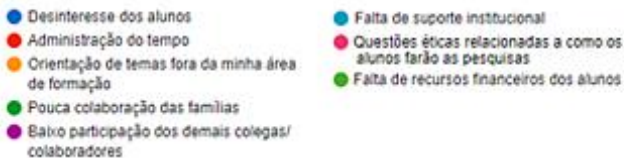

11 - Qual a 2ª maior dificuldade que pode surgir?

12 respostas

● Desinteresse dos alunos
● Administração do tempo
● Orientação de temas fora da minha área de formação
● Pouca colaboração das famílias
● Baixo participação dos demais colegas/ colaboradores
● Falta de suporte institucional
● Questões éticas relacionadas a como os alunos farão as pesquisas
● Falta de recursos financeiros dos alunos

Translation note:

11. What's the second biggest difficulty that can arise?

12 responses

Blue: Lack of interest from students
Red: Time management
Orange: Guidance on themes outside my area of training
Green: Little collaboration from families
Purple: Poor collaboration from other colleagues/peers
Light Blue: Lack of institutional support
Pink: Questions related to ethics regarding student participation in research
Light Green: Lack of financial resources of students

### Diversity of Themes and Selection for the Showcase

The themes chosen by the students reflected a wide diversity, and it is important to highlight that this did not imply a lack of availability from the lead teachers to provide guidance. The presence of the researcher in this school community was conceived as support meant to complement and assist the teachers in applying the PBL methodology. This approach demands not only IC guidance but also consistent follow-up, feedback provision, and support to the school community throughout the process.

Several research groups proposed cross-disciplinary themes beyond their advisors' areas of expertise (Table 1), with which most of the lead teachers expressed discomfort, as stated earlier.

**Table 1**

*Contrast between some of the research themes and the areas of expertise of the respective advisors*

| Theme | Subject Knowledge | Area | Field of Expertise |
| --- | --- | --- | --- |
| Femicide and Maria da Penha Law | Violence and Abuse | Sociology | Mathematics |
| The Boy Who Harnessed the Wind | Environment | Humanities | English Language |
| Women's Empowerment in Ballet | Social Awareness | Sociology | Portuguese Language |
| Belly Dancing and Women's Empowerment | Social Awareness | Sociology | Sciences |
| The Twin Paradox | Science and Technology | Physics | History |
| Violence Against Women | Violence and Abuse | Sociology | Mathematics |
| Sexual Harassment in Schools | Violence and Abuse | Sociology | Mathematics |
| Capitalist and Socialist Division in the Cold War | Human Development | Technology | Mathematics |

It is observed that the research themes chosen by the students reflected not only their personal interests but also the influence of the social and educational environment.

Out of the initial 102 workgroups, only 90 groups (88.2%) presented research projects. Among these, only 30 groups (33.3%) met the necessary criteria to be classified for the second part of the scientific exhibition, which would be presented to the community later.

However, this reduction from 102 to 90 research groups did not result in exclusions but in guiding students to integrate with research themes and groups of interest to them, aligned with the PBL methodology. This underscores the influence of social relationships and the teacher in learning. This integration stimulated the revision of themes and the formation of groups, allowing the reallocation of participants. It reinforces student autonomy and the pursuit of knowledge based on their interests without excluding the curriculum. Moreover, it is important to note that the other 60 groups that did not fully meet the evaluative rubrics followed all the steps of the proposed research initiation, aligning with the principles of the PBL methodology.

In light of this process, the new challenge was to keep students motivated in their investigations and to assist schoolteachers in meeting all this demand for research initiation guidance on various topics.

On the other hand, the initial resistance from teachers was partly fuelled by the underestimation of student abilities. However, as students began to show interest and satisfaction with the environment being created, and motivation became evident, the guiding teachers clearly saw that the students were fully capable of engaging in the process. It is important to highlight that, from the beginning, the school showed unwavering support, remaining available for teachers, and adopting a proactive approach to facing resistances.

The dynamic interaction between students, teachers, and ChatGPT as a co-guidance aligns intrinsically with Vygotsky's conception of cognitive development mediated by social interaction. While Vygotsky emphasized the importance of collaboration for knowledge construction, the research demonstrates how technological mediation can expand this interaction and enable an even richer and more diversified approach. This convergence between theory and results strengthens the idea that the conscious use of ChatGPT can be an effective catalyst for transforming education, without losing sight of the pedagogical foundations that underpin teaching practice.

Throughout the research process, the researcher was directly involved for approximately 65 hours of guidance and development with the workgroups, including activities of probing, planning, and pedagogical and methodological practices.

The researcher directly guided 54 research groups with the help of ChatGPT on various topics, revealing a dynamic that can be understood in light of Vygotsky's theory. In this context, ChatGPT acts as an "extended cultural mediator," expanding the ZPD of students and teachers. In general, the research themes chosen by students from the 90 groups were related to various knowledge areas, revealing varied interests and significant engagement. The analysis of themes, as illustrated in Figure 9, shows that 18 (20%) of the topics addressed issues related to "Violence and abuse," while 13 (14.4%) of them relate to "Health and quality of life". Additionally, 8 (8.9%) of the themes were linked to "Social awareness" and 7 (7.8%) addressed themes within "Science and technology".

**Figure 9**

*Diversity of thematic interests of the students from the 90 research groups.*

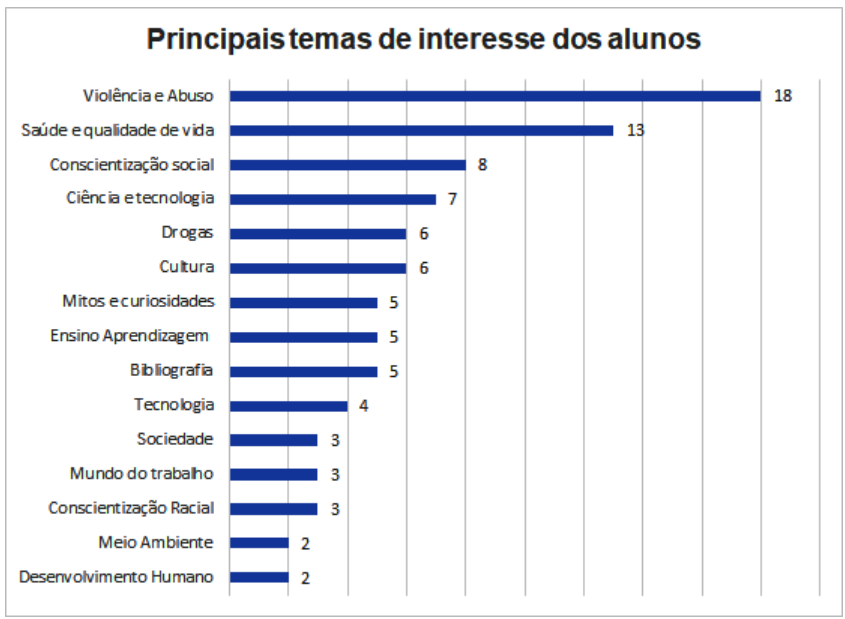

Translation note:

| Main topics of interest to students | Myths and curiosities |
|---|---|
| | Teaching and Learning |
| Violence and Abuse | Biography |
| Health and quality of life | Technology |
| Social awareness | Society |
| Science and technology | Working world |
| Drugs | Racial awareness |
| Culture | Environment |
| | Human Development |

Figure 9 reveals the different areas of interest of students concerning research themes. This distribution reflects not only the diversity of student interests but also the ability of the PBL methodology, with the support of ChatGPT, to stimulate the exploration of relevant and contemporary topics. Therefore, collaborative teaching practice integrated with active and transformative methodologies is essential during the teacher's mediation in the classroom.

This variety of themes required from the guides not only an interdisciplinary approach but also a sensitivity to cross-cutting issues, allowing teachers to promote the development of active citizenship and to build a responsible and inclusive social identity.

Notably, the enthusiastic participation of students in the SI projects and the tangible results achieved functioned as catalysts for the change in teachers' attitudes. Initially sceptical about innovative methodologies, educators were gradually witnessing a transformation in students' interest and attitude, which helped to overcome initial resistances. As the project progressed, the practice of SI was gaining support, becoming seen not only as a valuable learning opportunity for the students but also as a means of professional and personal development for the teachers themselves. These elements highlight the need to work on teachers' trust in their students and provide ongoing support in the process, validating the importance of social relations and teacher mediation in the teaching and learning process.

It is observed that students showed concern for a variety of themes such as "Drugs," "Science and Technology," "World of Work," "Health and Quality of Life," "Social Awareness," and "Racial Awareness." These themes are directly linked to citizenship and social identity, as well as interpersonal

__________________
Corresponding author: Renato P. dos Santos. Email: RenatoPSantos@ulbra.edu.br

relationships. The concern with health and quality of life reflects the students' interest in understanding and promoting physical, emotional, and social well-being. Social and racial awareness indicates a willingness to understand and contribute to relevant social issues, such as inequality and social injustice. These interests also demonstrate the maturity of the students and their willingness to engage in discussions and actions that are directly associated with their lives and society as a whole.

Analysing the relationships of interdisciplinarity, as shown in Figure 10, with the areas of knowledge, it can be seen that the research themes of the groups are predominantly related to the humanities, emphasizing sociology with themes of social, behavioural, and cultural involvement.

**Figure 10**

*Relationships of Interdisciplinarity in the research themes, highlighting the area of human sciences and themes related to sociology, social, behavioural, and cultural involvement.*

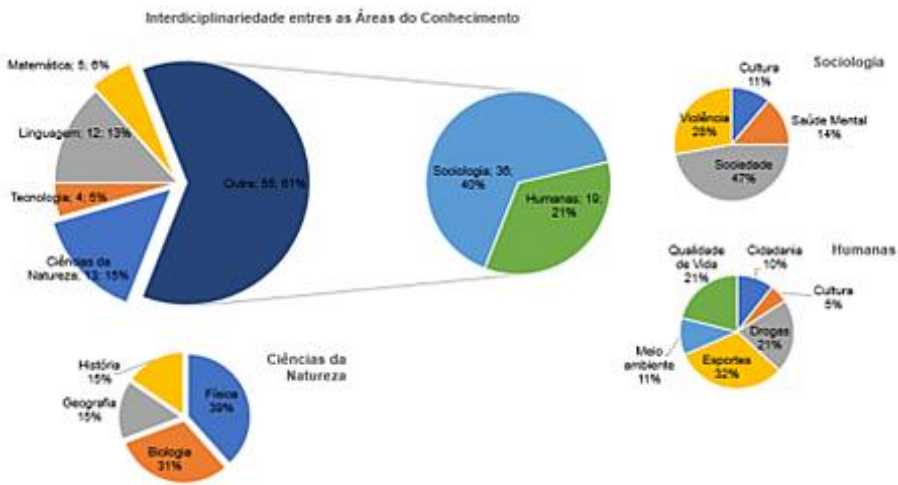

Translation note:

Interdisciplinarity among the Areas of Knowledge

| | |
|---|---|
| Mathematics –6% | Other –61% |
| Linguistics –13% | |
| Technology –5% | Sociology – 40% |
| Natural Sciences –15% | Other Human Sciences – 21% |

History – 15%

Geography – 15%

Biology – 31%

Physics – 30%

Quality of Life – 10%

Environment – 11%

Sports – 12%

Drugs – 21%

Culture – 5%

Citizenship – 10%

It is also possible to identify research themes related to the natural sciences area, which includes disciplines such as physics, biology, history, and geography. These themes explore natural phenomena, biological processes, historical events, and geographical aspects associated with technologies like the Cold War and space travel.

Given this perception, the importance of exchanging experiences, skills, and challenges among the faculty is evident. Sharing information and knowledge is essential for an interdisciplinary approach, which is also advocated in the PBL methodology. This collaboration contributes significantly to the development of any project, where the focus is on student protagonism in an active and transformative teaching practice.

Table 2 presents the themes, areas of knowledge, and advisors of the 30 groups that were selected by external evaluators at the Scientific Exhibition, which occurred in June 2023.

**Table 2**

*Themes, areas of knowledge, and advisor of the 30 groups selected for the June 2023 scientific exhibition.*

| Theme | Area of knowledge | Advisor |
|---|---|---|
| Belly Dance and Female Empowerment | Sociology | Researcher |
| The Disease of the Century: Anxiety | Sociology | Researcher |
| The Influence of Music | Language | Researcher |
| The Influence of Pythagoras on Mathematics Learning | Mathematics | Researcher |
| The Influence of Rapper Sabotage on the Current Brazilian Rap Scene | Sociology | Researcher |
| The Influence of Racionais MC's on Black Culture | Sociology | Researcher |
| Abuse and the Feelings It Can Bring | Sociology | Portuguese teacher |
| Attack in Brazilian Schools | Sociology | Sciences teacher |
| Calisthenics and Its Effects | Human Sciences | Phys. Ed teacher |
| Physical and Emotional Consequences of Artistic Skating on Female Bodies | Human Sciences | Researcher |
| Depression and Anxiety | Sociology | Researcher |
| Deforestation of the Amazon Rainforest | Human Sciences | Researcher |
| Devaluation of Women in Brazilian Sports | Human Sciences | Researcher |
| Fake News | Sociology | Researcher |
| Extreme Aesthetic Factors | Sociology | Researcher |
| Femicide at Home | Sociology | Math teacher |
| Suicide Rates in Rio Grande do Sul | Sociology | Math teacher |
| Influence of Math Teaching Channels | Mathematics | Researcher |
| Gender Inclusion in Sports | Sociology | Researcher |
| The Habit of Physical Activity Outside School | Human Sciences | Phys. Ed teacher |
| The Impact of Rapper Sabotage on Rap and Black Culture | Sociology | Arts teacher |
| The Mystery of the Bermuda Triangle | Natural Sciences | Portuguese teacher |
| Obesity in Adolescence | Human Sciences | Phys. Ed teacher |
| Cats | Natural Sciences | Sciences teacher |
| Society's Standard | Sociology | Researcher |
| Why Do Attacks Happen in Schools | Sociology | Sciences teacher |
| The Cost of Your Shower | Mathematics | Researcher |
| A Study on Minecraft | Technology | Sciences teacher |
| Truths and Lies About the Movie Interstellar | Natural Sciences | Researcher |
| Domestic Violence | Sociology | Sciences teacher |

It is worth noting the suggestive result that, of the 30 research groups qualified, 17 (56.67%) of them received direct guidance from the researcher, coupled with the co-guidance provided by ChatGPT (Table 2).

To investigate the possible influence of advising on the distinction of workgroups, a statistical analysis of the 90 participating groups was conducted. As mentioned before, of these 90 groups, 54 (60%) were advised by the researcher assisted by ChatGPT acting as a co-advisor, while the remaining 36 (40%) groups were advised by more experienced teachers. Of the 30 highlighted groups, 17 (31.5%) were among those 54 advised by the researcher with ChatGPT assistance, while 13 (36.1%) were among the 36 advised by various teachers.

The objective of the analysis was to determine whether the probability of a group being distinguished was associated with the type of advising received. Thus, the null hypothesis for the statistical test was that the chance of being distinguished would be independent of the advising, while the alternative hypothesis posited that there would be a difference in the proportions of distinguished groups between the two categories of advising.

The Fisher's exact test (Fisher, 1935) was used to evaluate the statistical significance of the association between the groups' advising and their distinction. The contingency table for the test was structured as in Table 3.

**Table 3**

*Contingency table showing the relationship between the type of group advising and their subsequent distinction.*

| Group | Distinguished | Not Distinguished | Totals |
|---|---|---|---|
| **Researcher + ChatGPT** | 17 (31,5%) | 37 (68,5%) | 54 (100%) |
| **Teachers** | 13 (36,1%) | 23 (63,9%) | 36 (100%) |
| **Totals** | 30 | 60 | 90 |

The test resulted in an odds ratio of 0.813, indicating a non-significant trend that the groups advised by the researcher with ChatGPT assistance were less likely to be distinguished compared to the groups advised by experienced teachers. The *p*-value obtained was 0.656, which exceeds the common threshold of 0.05 for statistical significance. Therefore, the data do not provide sufficient evidence to reject the null hypothesis of independence between the type of advising and the distinction of the groups.

These results suggest that, within the context of this study and the statistical power achieved, the probability of a group being distinguished was not significantly influenced by the advising profile, whether by the researcher assisted by ChatGPT or by the more experienced teachers.

Although the difference in the proportion of scientific initiation groups that were distinguished was not statistically significant, a promising trend was observed. Notably, 56.7% of the groups that were distinguished were those under the advising with the aid of ChatGPT. Furthermore, it is interesting to note that these 17 distinguished groups had the advising of a single teacher advising disparate themes, while the remaining distinguished groups were distributed among various teachers. This observation suggests a potential efficacy of using ChatGPT in the consistency and quality of the advising provided, pointing to the influence of unity in advising as a relevant factor for the success of scientific initiation projects.

This process of advising assisted by ChatGPT, including specific examples, will be detailed in the next subsection, providing a deeper understanding of this pedagogical approach.

### The Co-Advising Process with ChatGPT

*Initial Theme: English Culture (6th Grade)*

This research theme addressed aspects of English culture, emerging from an interdisciplinary collaboration that brought together the knowledge of a researcher (Physics teacher) and the expertise of an English teacher. This process is illustrated through a representative dialogue (Figure 11), where the integration of ChatGPT into the advising allowed the researcher to guide the student group to explore various elements of English-speaking countries' culture, such as cuisine, seasons, music, and the historical and political context, providing a comprehensive and multifaceted view.

**Figure 11**

*Dialogue during advising on the theme of English culture (6th Grade).*

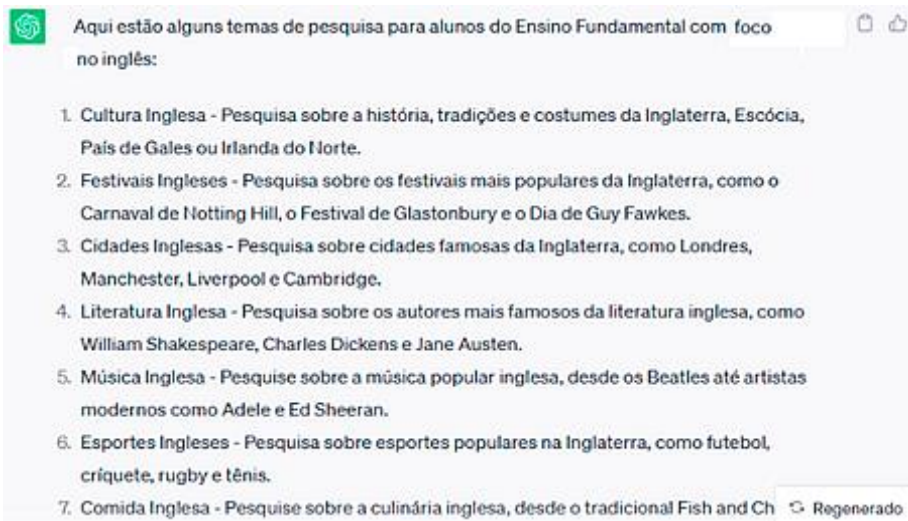

Translation note:

Here are some research topics for Elementary Education with a focus on English:

1. English Culture - Research the history, traditions, and customs of England, Scotland, Wales, and Northern Ireland.
2. English Festivals - Research about the most popular festivals in England, such as the Notting Hill Carnival, the Glastonbury Festival, and Guy Fawkes Day.
3. English Cities - Research famous cities in England, like London, Manchester, Liverpool, and Cambridge.
4. English Literature - Research about the most famous authors in English literature, such as William Shakespeare, Charles Dickens, and Jane Austen.
5. English Music - Research about popular music in England, from the Beatles to contemporary artists like Adele and Ed Sheeran.
6. English Sports - Research about popular sports in England, such as football, cricket, rugby, and tennis.
7. English Food - Research about culinary traditions, from the traditional Fish and Chips to regional dishes.

Although the English teacher did not interact directly with ChatGPT, her pedagogical practice was positively influenced by the collaborative dynamics and by the encouragement of cultural exploration within the scope of the research. The subsequent interdisciplinary study of English culture revealed the importance of embracing innovative approaches and the essential role of teacher mediation, which, even if not explicitly recognized, is crucial in promoting deep and engaged learning. The outcomes reflected the positive response of the students, evidenced by their satisfaction and the engagement demonstrated throughout the project.

The use of ChatGPT as an advising tool illustrates how technology can be used to expand the students' and teachers' ZPD, stimulating exploration and learning beyond their individual capabilities. The adaptation of the scientific method by the English teacher, inspired by the interdisciplinary collaboration, reveals a tacit understanding of the ZPD by balancing challenges and support to advance student development.

*Initial Theme: Belly Dance (7th Grade).*

Initially, a student, driven by an interest in belly dance, faced the challenge of defining the theme, identifying problems, and formulating concrete hypotheses. ChatGPT, as a digital co-advisor, played a crucial role by offering guidance that expanded both the advisor's and the group's understanding of the topic. Figure 12 illustrates this enriching dialogue.

**Figure 12**

*Exploration of Belly Dance - Challenges and Guidance.*

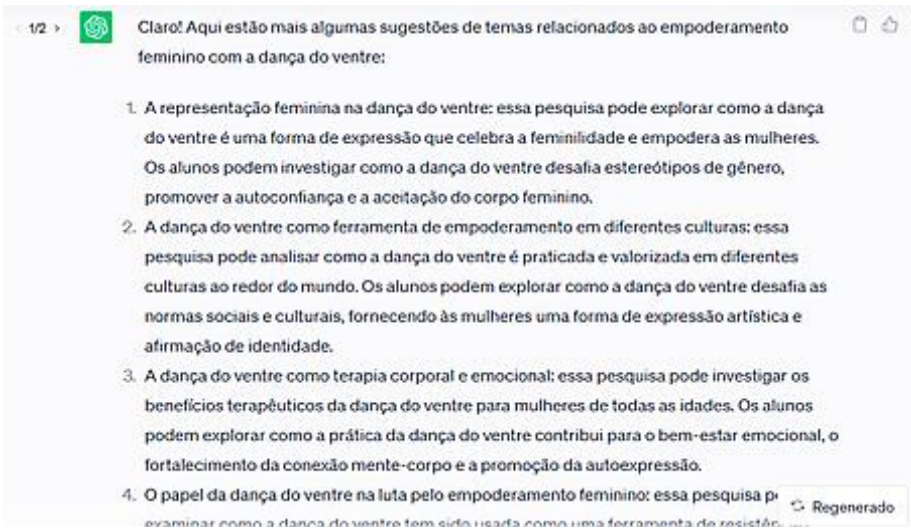

Translation note:

Hello! Here are some research topic suggestions related to women's empowerment through belly dance:
1. Female Representation in Belly Dancing: This research can explore how belly dancing is a form of expression that celebrates femininity and empowers women. Students can investigate how belly dance can be framed in gender studies and provide a counter-narrative to gender oppression.
2. Belly Dance as an Instrument of Body Empowerment in Different Cultures: This research can analyse how belly dance is perceived and valued in different cultures around the world. Students can explore how belly dance challenges social and cultural norms, offering women a form of artistic expression and identity affirmation.
3. Belly Dance as Body and Emotional Therapy: This research can investigate the therapeutic benefits of belly dancing for women of all ages. Students can explore how the practice of belly dance can contribute to emotional well-being, mental-body connection, and self-empowerment.

4. The Role of Belly Dance in Women's Empowerment: This research can examine the role of belly dance as an instrument for women's empowerment. The research can explore the connection between dance and personal development, body empowerment, and the promotion of well-being.

Co-advised by ChatGPT, the group discovered that belly dance could transcend its artistic nature, becoming a vehicle for female empowerment. Driven by this new perspective, both the advisor and the students embarked on exploring how belly dance could contribute to strengthening women's self-esteem, fostering a more positive relationship with their own bodies, and challenging the deep-rooted gender norms in society.

The research evolved to encompass intersectional issues of art, society, and gender, reflecting the synergy between the students, the teacher, and ChatGPT. This dynamic collaboration allowed for a significant expansion of their ZPDs, encouraging the students to explore topics that were beyond their prior knowledge, such as the connections between belly dance and female empowerment, as well as that of the teacher, since he was challenged to provide guidance on a subject far from his own training.

The student's statement, "I am very happy and proud, it was only with your guidance and that of the teacher that I managed to better understand what to do and then it was easy," highlights the importance of the received advising and underscores the effectiveness of intentional educational mediation. It expresses the essence of Vygotsky's ZPD: with the appropriate support, students can overcome challenges and elevate their understanding and skills. This case exemplifies how effective mediation can be crucial for meaningful learning on complex topics.

*Initial Topic: Amazon Deforestation (8th Grade)*

The students' investigation into Amazon deforestation began with research on global forest deforestation (Figure 13), which revealed that the Amazon, one of the world's largest forests, continues to face deforestation.

**Figure 13**

*Deforestation in the Amazon - Exploring the environmental impacts on the world's largest forest.*

qual a floresta que ocorre mais desmatamento no mundo?

A floresta que mais sofre com o desmatamento no mundo é a Floresta Amazônica, localizada principalmente no Brasil, mas também em outros países da América do Sul, como Peru, Colômbia, Bolívia, Equador, Guiana, Suriname, Venezuela e Guiana Francesa. A Amazônia é considerada uma das maiores florestas tropicais do mundo, abrangendo cerca de 6,7 milhões de milhas quadradas, e é fundamental para a biodiversidade global, regulando o clima e armazenando grandes volumes de carbono. Infelizmente, o desmatamento na Amazônia tem sido uma preocupação constante nas últimas décadas, principalmente devido à expansão da agricultura, pecuária e atividades madeireiras.

Translation note:

Which forest suffers the most deforestation in the world?

The forest that suffers the most from deforestation in the world is the Amazon Rainforest, primarily located in Brazil but also extending into other South American countries, such as Peru, Colombia, Bolivia, Ecuador, Guyana, Suriname, Venezuela, and French Guiana. The Amazon is considered one of the largest tropical forests in the world, encompassing about 67 million square miles. It is fundamental to global biodiversity, climate regulation, and the storage of large volumes of carbon. Unfortunately, deforestation in the Amazon has been a constant concern over the last few decades, mainly due to the expansion of agriculture, livestock farming, and mining activities.

Curiosity led the students to embark on a journey to understand and become aware of the environmental impacts and the relationship between deforestation and the Amazonian ecosystem, investigating deforestation rates over the past five years (Figure 14). This preliminary phase was crucial for establishing a factual foundation and awakening critical awareness among the students.

**Figure 14**

*Deforestation Rates in the Amazon Over the Last 5 Years - Exploring the increasing challenges and impacts in an interdisciplinary manner.*

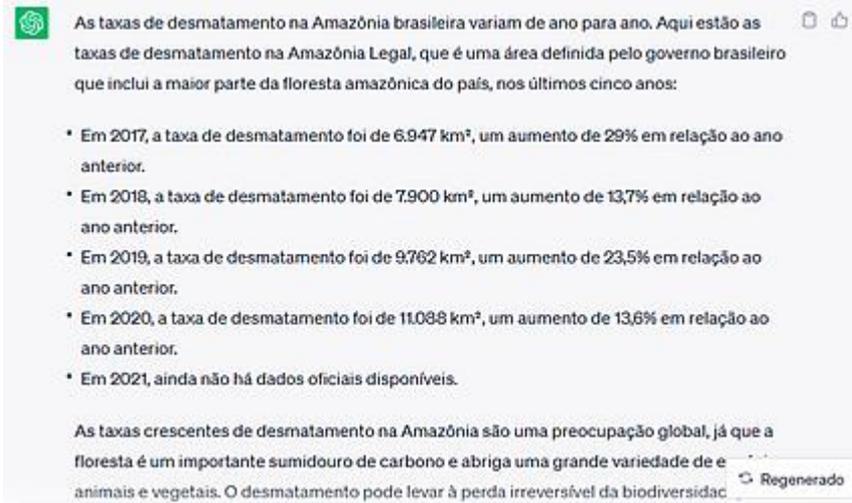

Translation note:

Deforestation rates in the Brazilian Amazon vary from year to year. Here are the deforestation rates in the Legal Amazon, which is a defined area by the Brazilian government that includes the largest part of the Amazonian Forest in the country, over the last five years:

- In 2017, the deforestation rate was 6947 km², an increase of 29% compared to the previous year.
- In 2018, the deforestation rate was 7900 km², an increase of 13.7% compared to the previous year.
- In 2019, the deforestation rate was 9762 km², an increase of 23.5% compared to the previous year.
- In 2020, the deforestation rate was 11088 km², an increase of 13.6% compared to the previous year.
- In 2021, data on deforestation is not yet officially available.

The increasing rates of deforestation in the Amazon are a global concern, as the forest is an important carbon sink and harbours a great variety of animals and plants. Deforestation could lead to irreversible biodiversity loss.

The experience of welcoming a new student to the research group highlighted the importance of sensitive and inclusive educational practices. His statement, "I joined the group later, but since I liked the topic it was easy to participate and introduce myself," demonstrates the ease with which this student engaged in the project and illustrates the positive impact of an open learning environment that is interested in the study topics.

Delving deeper into the subject, the students explored Brazilian environmental legislation, seeking to understand the nuances of Amazon conservation policies. The following dialogue with ChatGPT (Figure 15) provided additional insights and critical guidance that helped shape the students' perspective on the effectiveness of existing laws.

**Figure 15**

*Co-guidance and the search for relevant information for SI advising.*

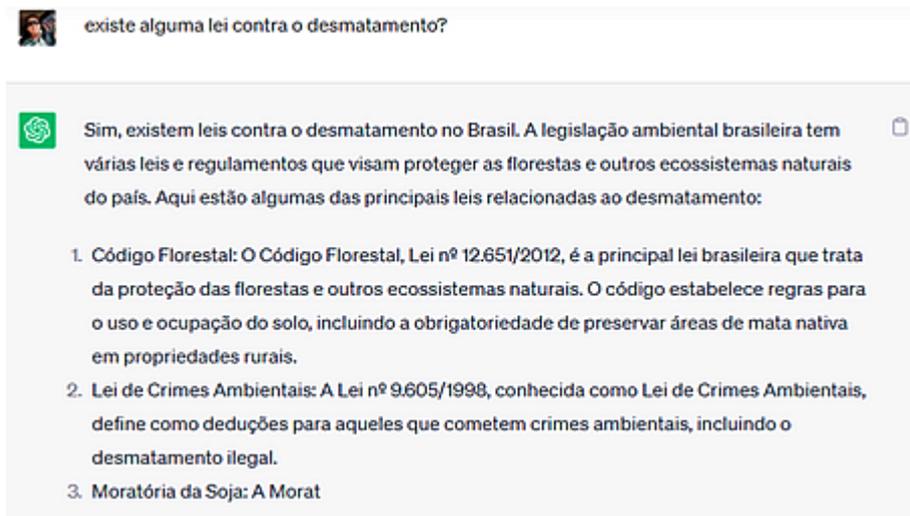

existe alguma lei contra o desmatamento?

Sim, existem leis contra o desmatamento no Brasil. A legislação ambiental brasileira tem várias leis e regulamentos que visam proteger as florestas e outros ecossistemas naturais do país. Aqui estão algumas das principais leis relacionadas ao desmatamento:

1. Código Florestal: O Código Florestal, Lei nº 12.651/2012, é a principal lei brasileira que trata da proteção das florestas e outros ecossistemas naturais. O código estabelece regras para o uso e ocupação do solo, incluindo a obrigatoriedade de preservar áreas de mata nativa em propriedades rurais.
2. Lei de Crimes Ambientais: A Lei nº 9.605/1998, conhecida como Lei de Crimes Ambientais, define como deduções para aqueles que cometem crimes ambientais, incluindo o desmatamento ilegal.
3. Moratória da Soja: A Morat

Translation note:
Yes, there are laws against deforestation in Brazil. The Brazilian environmental legislation includes various laws and regulations that aim to protect the forests and other natural ecosystems of the country. Here are some of the main laws related to deforestation:

1. Forest Code: The Forest Code, Law No. 12,651/2012, is a principal law that outlines the protection of forests and other natural ecosystems. The code establishes rules for the use and occupation of land, including the obligation to preserve areas of native vegetation on rural properties.

2. Environmental Crimes Law: The Law No. 9,605/1998, known as the Environmental Crimes Law, defines legal deductions for those who commit environmental crimes, including illegal deforestation.
3. Soy Moratorium: The Soy Morat [truncated]

As the students delved deeper into the topic (Figure 16), their ZPDs developed as they confronted the complexities of deforestation laws and how actions within and beyond legal boundaries affect the Amazon, considering the role of these laws in protecting the ecosystem. In this sense, the researcher sought to mediate learning, taking into account the cognitive development stage of the students, in line with Vygotsky's ideas.

**Figure 16**

*Deepening understanding of the complexity of deforestation laws in the Amazon.*

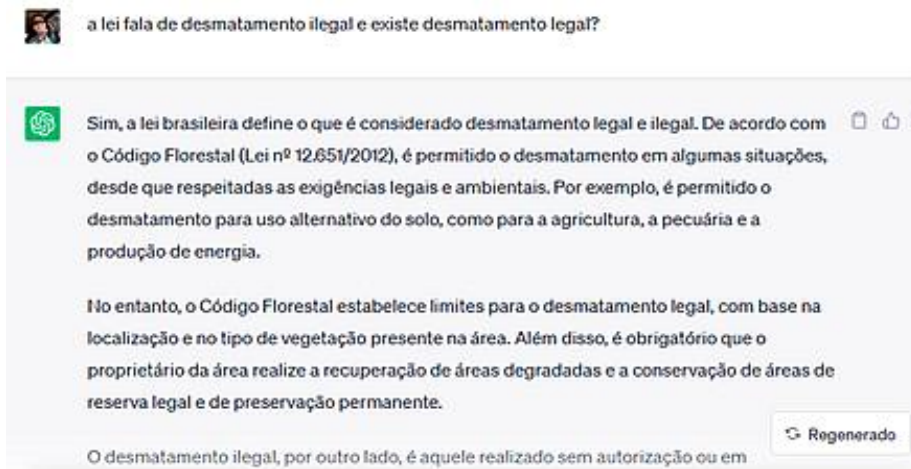

Translation note:

Yes, Brazilian law defines what is considered legal and illegal deforestation. According to the Forest Code (Law No. 12,651/2012), deforestation is allowed under some circumstances, provided that legal and environmental requirements are met. For example, deforestation for alternative land uses, such as for agriculture, livestock, and energy production, is permitted.

However, the Forest Code establishes limits for legal deforestation, based on the location and the type of vegetation present in the area. In addition, it is mandatory for the landowner to recover degraded areas and to conserve areas of permanent preservation and legal reserve.

Illegal deforestation, on the other hand, is carried out without authorization or in violation of legal provisions.

The educational process was enhanced using ChatGPT, which served as a resource to organize information and facilitate complex thinking. Figure 17 highlights the discussions that arose from this collaboration, encouraging the development of the students' critical thinking and a multifaceted analysis of the legal, social, and political issues involved in deforestation.

**Figure 17**

*Discussions on the complexity of deforestation.*

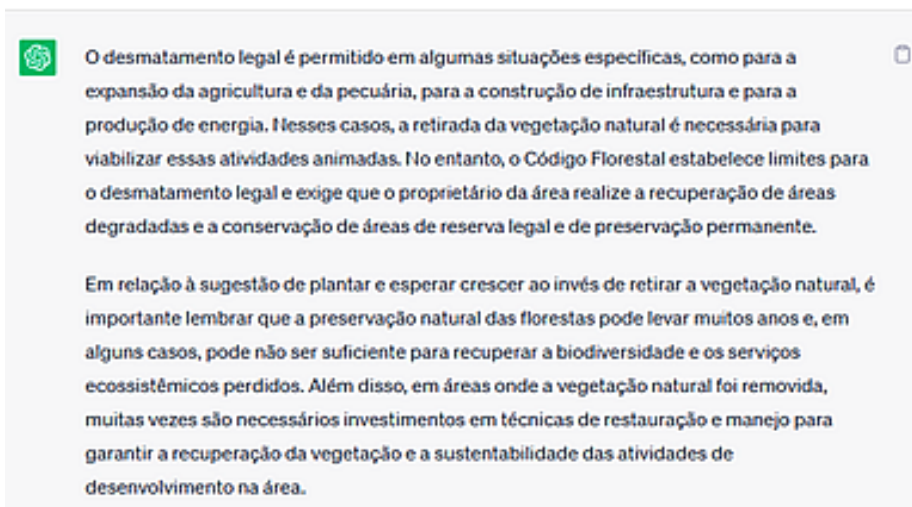

O desmatamento legal é permitido em algumas situações específicas, como para a expansão da agricultura e da pecuária, para a construção de infraestrutura e para a produção de energia. Nesses casos, a retirada da vegetação natural é necessária para viabilizar essas atividades animadas. No entanto, o Código Florestal estabelece limites para o desmatamento legal e exige que o proprietário da área realize a recuperação de áreas degradadas e a conservação de áreas de reserva legal e de preservação permanente.

Em relação à sugestão de plantar e esperar crescer ao invés de retirar a vegetação natural, é importante lembrar que a preservação natural das florestas pode levar muitos anos e, em alguns casos, pode não ser suficiente para recuperar a biodiversidade e os serviços ecossistêmicos perdidos. Além disso, em áreas onde a vegetação natural foi removida, muitas vezes são necessários investimentos em técnicas de restauração e manejo para garantir a recuperação da vegetação e a sustentabilidade das atividades de desenvolvimento na área.

Translation note:

Legal deforestation is permitted in certain specific situations, such as for the expansion of agriculture and livestock, for the construction of infrastructure, and for energy production. In these cases, the removal of natural vegetation is necessary to facilitate these economic activities. However, the Forest Code establishes limits for deforestation and requires that the landowner carry out the recovery of degraded areas and the conservation of areas of permanent preservation and legal reserve.

Regarding the suggestion of planting and waiting for new forests to replace natural vegetation, it is important to remember that the natural regeneration of forests can take many years and, in some cases, may not be sufficient to recover biodiversity and its ecosystem services within the necessary time frame. Also, when natural vegetation is removed, many times the soil suffers erosion, loss of organic matter, and nutrient leaching. Therefore, investments in areas where natural vegetation has been removed are essential, both to promote the recovery of vegetation and to ensure the sustainability of development in the area.

One of the questions raised by the students, "If it is so difficult and takes so long to plant and wait for growth, shouldn't deforestation just be illegal?" led to reflection by the teacher and a productive interaction with ChatGPT (Figure 18), encouraging the development of the students' critical thinking and their understanding of the complexities of deforestation.

**Figure 18**

*ChatGPT assisting in the organization of information and the direction of complex thought.*

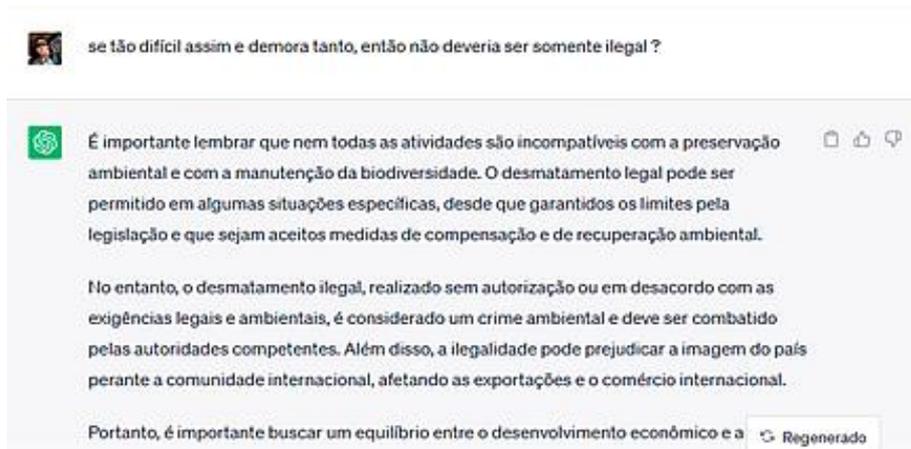

Translation note:

It is important to remember that not all soil-occupying activities are compatible with environmental preservation and biodiversity maintenance. Legal deforestation can be permitted in some specific situations, provided that the legal limits are respected and that there are measures for environmental compensation and recovery.

On the other hand, illegal deforestation, carried out without authorization or in disagreement with legal and environmental requirements, is considered an environmental crime and must be combated by the appropriate legal authorities. Additionally, illegal logging can damage the country's image in the international community, affecting exports and international trade.

Therefore, it is important to seek a balance between economic development and environmental conservation.

In conclusion, the adoption of the PBL methodology and ChatGPT as a guide in undergraduate research proved to be fundamental in expanding the students' ZPD. The teacher's mediation, aligned with the findings and challenges presented by the project, was crucial for guiding the students in their exploration of the complex legal and social aspects of deforestation, promoting meaningful and profound learning, aligned with the principles of Vygotsky's ZPD.

### Initial Topic: Gender Inclusion in MMA (9th Grade)

The research on gender inclusion in the sport of MMA began with a diversity of opinions among the students, highlighting the intrinsic complexity of the topic. The initial interaction, captured in Figure 19, shows the teacher's guidance, which, by fostering constructive debates, encouraged the students to expand their understanding of gender inclusion, aligned with Vygotsky's concept of ZPD.

As discussions deepened, the students, guided by interactions with ChatGPT (Figure 20), explored various facets of gender inclusion in MMA. This pedagogical approach encouraged critical questioning, enabling them to develop a more nuanced comprehension of the social implications and the issue of gender equality in the MMA sports context.

**Figure 19**

*Beginning of Research on Gender Inclusion in MMA*

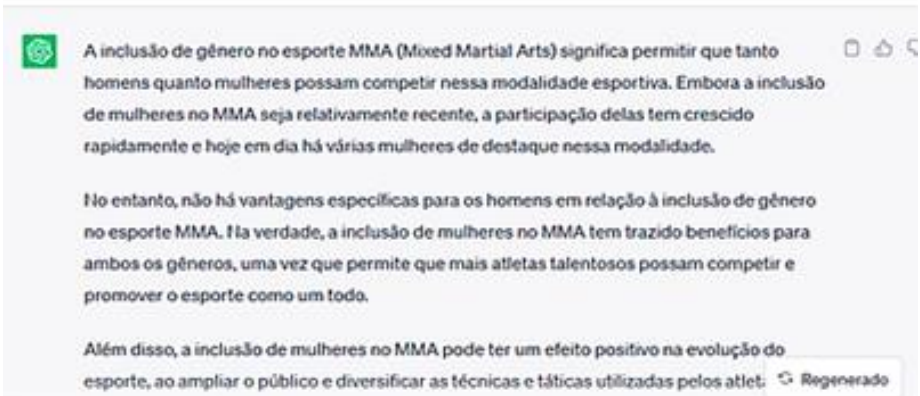

Translation note:

The inclusion of gender in the sport of MMA (Mixed Martial Arts) means that both men and women can equally participate in this sport. Although the involvement of women in MMA is relatively recent, their participation has grown rapidly and today there are several prominent women in this modality.

However, there are no specific advantages for men in relation to gender inclusion in MMA. In fact, the inclusion of women in MMA has been beneficial for the sport as a whole, as it allows for the most talented athletes to compete and promote the sport, something that everyone benefits from.

In addition, the inclusion of women in MMA can have a positive effect on the evolution of the sport, expanding and diversifying the techniques and tactics used by athletes.

**Figure 20**

*Dialogue guiding students to explore different approaches related to gender inclusion in the sport of MMA.*

como incluir esse assunto como tema de pesquisa?

Se você deseja incluir o tema de inclusão do gênero no esporte MMA como um tema de pesquisa, há várias abordagens que podem ser exploradas. Algumas sugestões de tópicos para pesquisa incluem:

1. A história da inclusão de mulheres no MMA: Explore como e por que as mulheres foram incluídas no esporte MMA e como essa inclusão evoluiu ao longo do tempo.
2. O impacto da inclusão de gênero no esporte MMA: Investigar como a inclusão de mulheres no esporte MMA afetou a popularidade, audiência e patrocínio do esporte, bem como a evolução técnica e tática das lutas.
3. A igualdade de gênero no esporte MMA: Analisar a igualdade de oportunidades entre homens e mulheres no esporte MMA, incluindo o acesso a treinamento, competições, patrocínios e pagamentos.
4. As diferenças entre as lutas de homens e mulheres no MMA: Investigar as diferenças nas habilidades técnicas e táticas de homens e mulheres no esporte MMA e como isso pode afetar a maneira como as lutas são disputadas.
5. A percepção pública da inclusão de gênero no esporte MMA: Explorar como o público em geral percebe a inclusão de mulheres no esporte MMA e como isso pode afetar a popularidade e o sucesso do esporte.
6. Incluindo

Translation note:

If you wish to include the theme of gender inclusion in the sport of MMA as a research topic, there are various approaches that can be explored. Some suggested topics for inclusion are:

1. The history of the inclusion of women in MMA: Explore how and why women were included in the sport of MMA and how this inclusion has evolved over time.
2. The impact of gender inclusion on the sport of MMA: Investigate how the inclusion of women in the sport of MMA affects its popularity, audience, and sponsorship of the sport, as well as the technical and tactical evolution of the sport.
3. Gender equality in the sport of MMA: Analyse the equality of opportunities between men and women in the sport of MMA, including access to training, competitions, and sponsorship deals.

4. The differences in techniques and tactics between men and women in MMA: Investigate the differences in strategies and styles between men and women in the sport of MMA and how this can affect the way fights are fought and the outcomes of matches.
5. Public perception of the inclusion of gender in the sport of MMA: Explore how the public perceives the inclusion of women in the sport of MMA and how this perception can affect the popularity and success of the sport.
6. Including [truncated]

In analysing physical differences and the role of hormonal therapy, students were guided to understand the complexity of biological issues related to gender inclusion in the sport of MMA. The guidance provided in the dialogue (Figure 21) allowed the students to explore topics in a sensitive and fact-based manner and reflect on the evolution of the physical characteristics of trans athletes and the challenges they face in the competitive context.

**Figure 21**

*Dialogue about the physical differences between genders and the influences of hormonal therapy on gender inclusion in MMA.*

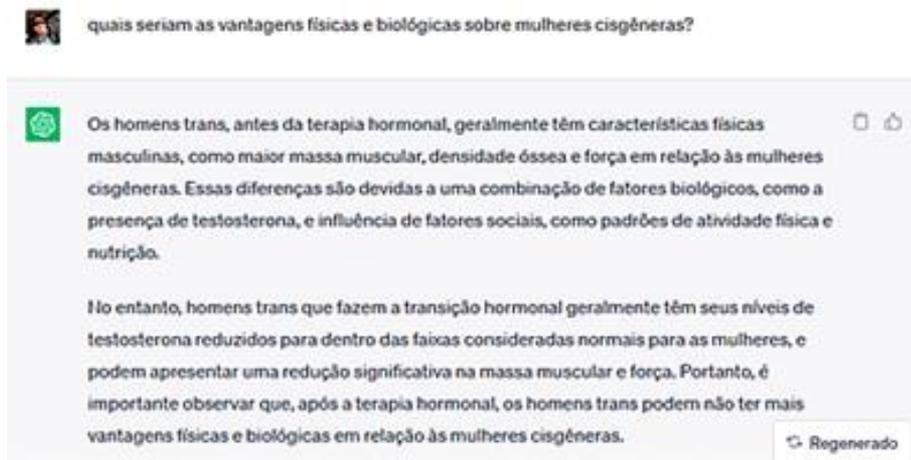

Translation note:

Trans men, after hormonal therapy, generally have physical characteristics that are more masculine, such as increased muscle mass, bone density, and strength in relation to cisgender women. These differences are due to a combination of biological factors, such as the presence of testosterone, and the influence of social factors, such as exercise patterns and nutrition.

However, trans men who undergo hormonal transition usually have their testosterone levels reduced to within ranges considered normal for women, and they may observe a significant reduction in muscle mass and strength. Therefore, it is important to note that, after hormonal therapy, trans men may not have more physical and biological advantages in relation to cisgender women.

**Figure 22**

*Dialogue exploring the lack of evidence of advantages of trans women over cisgender men in athletic or physical abilities.*

como pode uma mulher trans ter vantagens sobre o homem cisgenero?

Não há evidências científicas que sugiram que mulheres trans tenham vantagens físicas sobre homens cisgêneros em esportes competitivos. Mulheres trans são pessoas que foram designadas como masculinas ao nascer, mas que se identificam como mulheres. Portanto, a maioria das mulheres trans pode ter níveis mais baixos de testosterona do que homens cisgêneros, o que pode resultar em menor força muscular e densidade óssea.

No entanto, é importante reconhecer que as diferenças biológicas entre homens e mulheres são complexas e variáveis, e podem ser influenciadas por fatores como idade, treinamento esportivo, nutrição e genética. Além disso, a vantagem em esportes competitivos não se resume apenas a diferenças biológicas, mas também inclui fatores como treinamento, habilidade, estratégia e motivação.

É importante garantir que todas as pessoas tenham o direito de participar de esportes competitivos, independentemente de sua identidade de gênero, desde que sejam respeitadas as regras e regulamentos estabelecidos pelas organizações esportivas. A inclusão de mulheres trans no esporte é um tópico controverso e em evolução, e as organizações esportivas estão trabalhando para desenvolver diretrizes claras e justas para garantir que todos os atletas possam competir em um ambiente seguro e equitativo.

Translation note:

How can a trans woman have advantages over a cisgender man?

There is no scientific evidence to suggest that trans women have physical advantages over cisgender men in competitive sports. Trans women are individuals who were designated male at birth but identify as women.

Therefore, the majority of trans women can have lower muscle mass and testosterone levels than cisgender men, which could result in less force due to the decreased muscle density.

Nevertheless, it is important to recognize that the biological differences between men and women are complex and variable, and can be influenced by factors such as age, training, sports nutrition, and genetics. In addition, the advantage in competitive sports is not solely reduced to biological differences but also includes factors such as training, skill, strategy, and motivation.

It is essential to ensure that all individuals have the right to participate in competitive sports, regardless of their gender identity, provided that they comply with the rules and regulations established by sports organizations. The inclusion of trans women in sports is a controversial and evolving topic, and sports organizations are working to develop fair and equitable criteria to ensure that all athletes can compete in a safe and equitable environment.

**Figure 23**

*Dialogue on the importance of reflection and ongoing commitment to promoting gender equality in sports.*

Quanto tempo vai demorar para que a inclusão do gênero seja aceita pela sociedade?

Não há uma resposta simples para essa pergunta, pois a inclusão de gênero pela sociedade depende de diversos fatores, como cultura, tradições, valores e política. No entanto, é importante destacar que a luta pela inclusão do gênero é um processo contínuo e que já houve avanços em várias partes do mundo.

No entanto, ainda há muitos desafios a serem enfrentados. Algumas sociedades ainda são profundamente conservadoras e resistem à mudança, e muitas vezes há resistência política à inclusão de gênero. Além disso, a discriminação de gênero ainda é uma realidade para muitas pessoas, especialmente mulheres e pessoas trans.

Acredita-se que a inclusão do gênero seja um processo gradual e que deve ser impulsionado por uma série de mudanças sociais e políticas. À medida que a conscientização sobre a importância da igualdade de gênero aumenta e à medida que as pessoas se tornam mais engajadas na luta pelos direitos das minorias de gênero, é possível que a inclusão de gênero seja mais amplamente aceita pela sociedade. No entanto, não há como prever com precisão quanto tempo isso pode levar.

Translation note:

How long will it take for gender inclusion to be accepted by society?

There is no simple answer to this question, as the acceptance of gender inclusion within society depends on a multitude of factors, including culture, traditions, values, and politics. However, it is essential to recognize that the push for gender inclusion is a continuous process that has already made significant progress in various parts of the world.

Nevertheless, there are still many hurdles to overcome. Some societies remain deeply conservative and resist change, sometimes even showing political resistance to gender inclusion. Moreover, discrimination based on gender is still a reality for many people, especially women and transgender individuals.

It is believed that the inclusion of gender in society is a gradual process that must be driven by a series of educational and policy changes. As awareness of the importance of gender equality grows and as individuals become more engaged in the fight for minority gender rights, it is possible that gender inclusion will be more widely accepted by society. However, it is uncertain how long it may take for this inclusion to be fully realized.

Issues regarding the lack of evidence of competitive advantages of trans women over cisgender men in athletic or physical abilities were heatedly debated in the group. The dialogue with ChatGPT (Figure 22) highlighted that gender identity does not directly determine athletic ability, emphasizing the importance of an evidence-based approach and its limits. By mentioning the measures taken by sports organizations to create fair guidelines, students were directed to reflect on the need for an equitable and well-founded approach.

Figure 23 illustrates the evolution of the research, which went from searching for quick answers to a more in-depth reflection on the cultural, political, and social aspects that shape gender inclusion in MMA, leading students to broaden their perspectives. Students were encouraged to consider the long road to social acceptance and the importance of continuous commitment to the cause of gender equality, which became the focus of the developed research.

The ongoing dialogue with ChatGPT helped to mediate learning, guiding the students through a process where questions and internal conflicts were overcome. As the students interacted with ChatGPT and received guidance from the teacher, they reflected on their understandings and expanded their ZPDs, engaging deeply with the topic and questioning prior beliefs, organically and collaboratively transitioning from an immediate search for answers to a richer and more mature understanding of the scientific method.

During this educational journey about gender inclusion in sports, the group experienced a transformational learning process where they moved from searching for definitive answers to valuing investigation, questioning, and gradual understanding. This statement from one of the group's students encapsulates this journey of discovery:

> *"Throughout the research, our group faced conflicts, and with guidance, we understood that in research we don't need to have an answer as an absolute truth, the opinion of some in the group changed over the course of the research, we thought about giving up, but we understood that this is how it is..."*

This approach demonstrated how proper teacher mediation, combined with ChatGPT technology, can provide an environment where intellectual curiosity flourishes and the learning potential of students expands, illustrating the dynamics of research and the transformation it can instigate in students.

## CONCLUSIONS

The findings of this research highlight resonance with the principles of Vygotsky's sociocultural theory, particularly regarding the incorporation of ChatGPT as a pedagogical tool in the current educational landscape. The thematic diversity chosen by the students revealed an expansion of their horizons, demonstrating the applicability of the Zone of Proximal Development (ZPD) and solidifying ChatGPT as a "co-advisor." The advisor, in turn, assumed a decisive "pedagogical mediator" role, fostering a collaborative learning environment and providing precise guidance.

Vygotsky argues that cognitive development is strongly shaped by social and cultural interactions, dismissing purely biological or innate interpretations. He emphasizes the influence of language, culture, and social tools in the shaping of human thought (Vygotsky, 1991). Modernity has brought a new digital culture, challenging educators to renew their pedagogical approaches with intentionality and innovation.

Thus, while Santos et al. (2020) pointed to the scarcity of time as an obstacle to the implementation of active methodologies, this research demonstrates the use of ChatGPT as a viable solution to optimize group advising time and to boost innovative methods in research guidance within the educational context. The integration of AI has proven not only efficient but also enriching, allowing the exploration of interdisciplinary and current themes, surpassing the temporal limitations identified by Santos et al. (2020). By using ChatGPT to structure information and guide workgroups, the advisor overcame resistances and lack of experience, consolidating technology as an essential ally in the mediation of learning.

It is essential to emphasize that the co-advisership provided by ChatGPT offered the teacher the ability to provide significant support, even in areas of knowledge that transcend the academic training of the researcher, enhancing the autonomy of students in their research. This methodology surpassed the barriers of advisors' specializations and reinforced the emerging position of AI as a valuable pedagogical resource. The use of ChatGPT demonstrated how technology could complement human expertise in the advising process and enhance students' learning opportunities. This innovative approach resulted in richer teaching practices and a deeper reflection on pedagogy, strengthening and consolidating the educational identity of the researcher.

In this study, a Physics teacher in teacher training, in his debut as a advisor in Scientific Initiation at the Elementary level, exceeded expectations and personal resistances, perceiving a transformation in teaching practice, with a clear social intentionality and commitment to scientific dissemination.

The findings also point to a remarkable synergy between theory and practice in the educational environment, integrating active methodologies, use of technology, and pedagogical action. They reflect the understanding that Education, more than a mere application of theories, is a complex and interactive practice that must be in constant evolution to enrich the student learning experience.

The study culminated in a notable cultural transformation within the school. This change, from a traditional view to a socialized learning approach, was catalysed by a proactive and non-imposing school management. The administration played a decisive role by delegating responsibilities and incorporating the community into the so-called "pedagogical stops," encouraging closer collaboration between students and teachers. Teachers' resistance to innovation, initially robust, was gradually overcome by the visible impact of the enthusiasm and active participation of the students in the Project-Based Learning process.

Notably, although ChatGPT was not directly used by students due to teachers' resistance, the movement towards pedagogical innovation positively influenced the school community. Student groups from the school gained recognition at prestigious events: one group won first place at Mostratec and another received an honourable mention at Femositec. These successes are particularly notable because they occurred in projects that were not under the direct guidance of the researcher. Such achievements demonstrate the indirect impact of the implemented methodology, evidencing the awakening of intrinsic motivation among students and the subsequent decrease in opposition from teachers.

Therefore, it is noted that, despite the initial resistance of teachers to adopting innovative methodologies and the use of artificial intelligence, the progress of the project and the visible enthusiasm of the students were crucial in overcoming these obstacles. The experience reaffirms the effectiveness of Project-Based Learning (PBL) in engaging students and teachers in a joint educational experience, promoting the socialization of knowledge and the renewal of pedagogical practices, which was proven by the tangible recognition through awards at local scientific fairs.

This study also exemplifies the effectiveness of action research in educational contexts, demonstrating how intentional actions and critical reflections can trigger significant cultural transformations, whose effects reverberate beyond the physical boundaries of the educational institution. The observed change in the school culture, transitioning from traditional practices to a collaborative and socialized approach, corroborates the ideals of action research. The active participation and engagement of teachers, students, and school management were reflected in the continuous improvement of pedagogical practices and the promotion of meaningful learning. The iterative cycles of planning and reflection facilitated the overcoming of initial resistances, culminating in external recognition through awards at scientific fairs. The importance of effective school management is highlighted, which, through its supportive and empowering role, significantly influenced the pedagogical renewal.

This study also exemplifies the effectiveness of action research in educational contexts, showing how intentional actions and critical reflections can trigger significant cultural transformations, whose effects reverberate beyond the physical boundaries of the educational institution. The change observed in the school culture, moving from traditional practices to a collaborative and socialized approach, corroborates the ideals of action research. The active participation and engagement of teachers, students, and school management reflected in the continuous improvement of pedagogical practices and the promotion of meaningful learning. The iterative cycles of planning and reflection facilitated the overcoming of initial resistances, culminating in external recognition through awards at scientific fairs. The importance of effective school management is highlighted, which, by providing support and delegating responsibilities, contributed decisively to the socialization of knowledge and the empowerment of participants.

In light of the promising results of this study, a fertile ground is envisioned for future investigations that can expand the understanding of the integration of active methodologies and emerging technologies in the educational context. It would be enriching to explore the sustained use of ChatGPT and other artificial intelligence tools across different disciplines and educational levels, assessing long-term impacts on student motivation and research skill development. Furthermore, longitudinal studies could investigate changes in teachers' attitudes towards innovative approaches and the role of school management in sustaining pedagogical transformations. Extending this work to other educational contexts, encompassing a wider range of sociocultural variables, could also provide valuable insights into the replicability and adaptability of project-based learning and action research methods. The interface between educational research and AI technologies remains a vibrant and underexplored frontier, promising significant advances in the field of education.

## ACKNOWLEDGEMENTS

FB gratefully acknowledges the partial funding from Programa Institucional de Bolsas de Iniciação Científica da Ulbra (PROICT/Ulbra), which was instrumental to this research.

## AUTHORS' CONTRIBUTION STATEMENT

FB was responsible for the conception and execution of this research, as well as the preliminary data collection and analysis. He also took charge of writing the first draft of the text and the review of the final version. RPdS was the research advisor. He also contributed to the review of the first version and the writing of the final version.

## DATA AVAILABILITY STATEMENT

The authors agree to make their data available upon reasonable request. It is up to the authors to determine whether a request is reasonable or not.

# APÊNDICE A

## Questionário aos professores e equipe diretiva

Respostas abertas ate o dia 20.04

Apresentação

A Universidade Ulbra busca aprimorar a prática pedagógica dos estudantes de licenciatura em formação, por meio do Programa de Extensão Interdisciplinar (PEI), em parceria com escolas e a sociedade, esse programa tem como objetivo promover uma abordagem mais ampla e colaborativa na formação de futuros profissionais da educação. Nosso projeto busca investigar

**"A importância da aprendizagem baseada em projetos no ensino fundamental".**

Desde já agradecemos a disponibilidade da da equipe diretiva e corpo docente da E.M.E.F. Irmão Pedro.

1. Questionário destinado a professores do ensino fundamental dos **anos iniciais, finais** * **e equipe diretiva/gestão**. Importante a participação de todos da escola, mesmo não envolvidos diretamente com o projeto.

   1 - Qual sua área de atuação na escola?

   *Mark only one oval.*

   ◯ Anos iniciais (1°- 5° ano)

   ◯ Anos finais (6°- 9° ano)

   ◯ Equipe diretiva/gestores (coordenação, supervisão, orientação)

2. 2- Qual a sua formação? *

*Mark only one oval.*

◯ Letras

◯ Educação Física

◯ Ciências Biológicas

◯ Filosofia

◯ Sociologia

◯ História

◯ Artes Visuais

◯ Geografia

◯ Matemática

◯ Física

◯ Ensino religioso

◯ Masgisterio/ pedagogia

◯ Outros

3. 3 - Qual das metodologias abaixo mais simpatiza? *

*Mark only one oval.*

◯ O método de ensino tradicional

◯ O método de ensino construtivista

◯ O método de ensino montessoriano

◯ O método de ensino waldorfiano

◯ O método de ensino sócio-interacionista

◯ Metodologias ativas

◯ Metodologias STEAM

◯ Aprendizagem baseada em projetos

4. 4 - Qual das teorias de ensino e aprendizagem, mais utiliza em sala de aula ou simpatiza? *

*Mark only one oval.*

◯ Teoria Experiencial

◯ Teoria cognitivista

◯ Teoria Social Cognitiva

◯ Teoria humanista

◯ Aprendizagem significativa

◯ Teoria sociocultural (da aprendizagem social, construtivista)

◯ Teoria do comportamento(Behaviorismo)

5. 5 - Conforme as práticas e metodologias de ensino e aprendizagem que utiliza *
abaixo, qual grau de envolvimento/participação dos seus alunos em sala de aula, de
modo geral?

*1 pouco envolvimento, 5 muito envolvimento*

*Mark only one oval per row.*

|  | 1 | 2 | 3 | 4 | 5 | não utilizo/não estou em sala |
|---|---|---|---|---|---|---|
| Aprendizagem baseada em projetos | ○ | ○ | ○ | ○ | ○ | ○ |
| Ensino tradicional | ○ | ○ | ○ | ○ | ○ | ○ |
| Ensino sócio-interacionista | ○ | ○ | ○ | ○ | ○ | ○ |
| Ensino construtivista | ○ | ○ | ○ | ○ | ○ | ○ |
| Estudo de caso | ○ | ○ | ○ | ○ | ○ | ○ |
| Mapas conceituais | ○ | ○ | ○ | ○ | ○ | ○ |
| Seminários e discussões | ○ | ○ | ○ | ○ | ○ | ○ |
| Trabalhos em grupos | ○ | ○ | ○ | ○ | ○ | ○ |
| Tecnologias de informação e comunicação | ○ | ○ | ○ | ○ | ○ | ○ |
| Prática multidisciplinar | ○ | ○ | ○ | ○ | ○ | ○ |

6. 6 - Referente as opção abaixo como percebe, de modo geral, o desenvolvimento das * turmas que leciona? *1 para ruim, 6 para muito bom*

*Mark only one oval per row.*

|  | 1 | 2 | 3 | 4 | 5 | 6 | não estou em sala |
|---|---|---|---|---|---|---|---|
| **Raciocínio lógico** | ◯ | ◯ | ◯ | ◯ | ◯ | ◯ | ◯ |
| **Investigação** | ◯ | ◯ | ◯ | ◯ | ◯ | ◯ | ◯ |
| **Comunicação** | ◯ | ◯ | ◯ | ◯ | ◯ | ◯ | ◯ |
| **Participação** | ◯ | ◯ | ◯ | ◯ | ◯ | ◯ | ◯ |
| **Compreensão** | ◯ | ◯ | ◯ | ◯ | ◯ | ◯ | ◯ |
| **Criatividade** | ◯ | ◯ | ◯ | ◯ | ◯ | ◯ | ◯ |

7. 7 - De modo geral, como você professor(a), educador(a) está? *
*1 pouco - 5 muito*

*Mark only one oval per row.*

|  | 1 | 2 | 3 | 4 | 5 |
|---|---|---|---|---|---|
| **Ansioso(a)** | ◯ | ◯ | ◯ | ◯ | ◯ |
| **Criativo(a)** | ◯ | ◯ | ◯ | ◯ | ◯ |
| **Cauteloso(a)** | ◯ | ◯ | ◯ | ◯ | ◯ |
| **Empático(a)** | ◯ | ◯ | ◯ | ◯ | ◯ |
| **Irritado(a)** | ◯ | ◯ | ◯ | ◯ | ◯ |
| **Inspirado(a)** | ◯ | ◯ | ◯ | ◯ | ◯ |
| **Preocupado(a)** | ◯ | ◯ | ◯ | ◯ | ◯ |
| **Resiliênte** | ◯ | ◯ | ◯ | ◯ | ◯ |

8. 8 - Já trabalhou com aprendizagem baseada em projetos, na escola Irmão Pedro? *

*Mark only one oval.*

◯ Sim

◯ Não

9. 9 - Já orientou alunos em iniciação científica, na escola Irmão Pedro ou em outra escola? *

*Mark only one oval.*

◯ Sim

◯ Não

10. 10 - Qual é a maior dificuldade que você acredita que pode ser encontrado no processo de orientação para iniciação científica que estamos iniciando? *

*Mark only one oval.*

◯ Desinteresse dos alunos

◯ Administração do tempo

◯ Orientação de temas fora da minha área de formação

◯ Pouca colaboração das famílias

◯ Baixo participação dos demais colegas/colaboradores

◯ Falta de suporte institucional

◯ Questões éticas relacionadas a como os alunos farão as pesquisas

◯ Falta de recursos financeiros dos alunos

11. 11 - Qual a 2° maior dificuldade que pode surgir? *

*Mark only one oval.*

◯ Desinteresse dos alunos

◯ Administração do tempo

◯ Orientação de temas fora da minha área de formação

◯ Pouca colaboração das famílias

◯ Baixo participação dos demais colegas/colaboradores

◯ Falta de suporte institucional

◯ Questões éticas relacionadas a como os alunos farão as pesquisas

◯ Falta de recursos financeiros dos alunos

12. 12 - Qual é o maior desafio que você espera enfrentar durante este processo? *
(Profs. dos anos iniciais, considerem aplicando essa prática)

*Mark only one oval.*

◯ Engajar e manter os alunos motivados

◯ Administração do tempo

◯ Compartilhar ou delegar dificuldades com temas fora da minha área

◯ Gerenciar conflitos internos de aluno por transtorno ou dificuldades de aprendizagem

◯ Orientar os grupos de pesquisa e conseguir ajudar outras turmas

◯ Buscar contextualizar a disciplina com temas diversos

◯ Dificuldades com uso de tecnologias digitais

◯ Não creio que a aplicação desta metodologia, nesse momento, seja a ideal

◯ Aplicar a metodologia em si de forma eficiente

**APÊNDICE B**

## Questionário com alunos anos finais

Iniciação científica escola Irmão Pedro

1. 1 - Qual sua série? *

   *Mark only one oval.*

   ⬭ a) 6° ano

   ⬭ b) 7° ano

   ⬭ c) 8° ano

   ⬭ d) 9° ano

2. 2 - Quando tu quer saber algo ou tirar alguma dúvida, qual seu principal meio *
   de pesquisa?

   *Mark only one oval.*

   ⬭ a) Normalmente pergunto aos meus pais

   ⬭ b) Normalmente pergunto aos meus profs

   ⬭ c) Pesquiso na Wikipédia

   ⬭ d) Pesquiso sempre no google

   ⬭ e) Pesquiso sempre no youtube

   ⬭ f) Livros, revistas

3.   3 - Acha mais importante saber: as fontes da pesquisa ou as referências? *

*Mark only one oval.*

⬭ a) Considero as Fontes importantes

⬭ b) Considero as Referências importantes

⬭ c) Desde que seja relevante e coerente com minha opinião, aceito!

⬭ d) Nunca pensei nisso, eu só pesquiso!

⬭ e) Não sei exatamente, mas acho que os dois

4.   4 - Iniciação científica, está relacionado com: *

*Mark only one oval.*

⬭ a) Em ser cientista

⬭ b) Em pesquisador

⬭ c) Em fazer experimentos

⬭ d) Em resolver problemas

⬭ e) Em saber questionar

⬭ f) Em saber perguntar e responder, com evidências

5.   5 - Como se sente aqui e agora, entes de iniciar a jornada de Iniciação           *
     Cientifica?

*Mark only one oval.*

⬭ a) Normal

⬭ b) Nervoso(a)

⬭ c) Não sei

⬭ d) Adoro coisas novas

⬭ e) Preocupado (a)

⬭ f) Chateado, mais uma coisa p fazer!

⬭ g) Ansioso(a)

⬭ h) Motivado (a)

⬭ i) Prefiro nao dizer